\documentclass[aps,prxquantum,reprint,twocolumn,superscriptaddress,floatfix,nofootinbib,longbibliography]{revtex4-1}
\usepackage{graphicx,amsmath,amsfonts,amssymb,amsthm,xr}
\usepackage{epsfig,amsmath,amssymb,color,dsfont,upgreek,physics}
\usepackage{mathrsfs}
\usepackage{mathtools}
\usepackage{bbold}
\usepackage{comment}
\usepackage{float}

\usepackage[bookmarks=true,colorlinks,linkcolor=OrangeRed,urlcolor=NavyBlue,citecolor=RoyalBlue]{hyperref}
\usepackage[dvipsnames]{xcolor}
\usepackage{orcidlink}

\definecolor{mygold}{rgb}{0.93,0.69,0.13}

\definecolor{mypurple}{rgb}{0.49,0.18,0.56}

\allowdisplaybreaks

\makeatletter
\newcommand\emailx[1]{%
\move@AF%
\def\@affil{{\normalfont\,#1\strut}{}}%
}%

\def\@fnsymbol#1{\ensuremath{\ifcase#1\or \dagger\or \ddagger\or
   \mathsection\or \mathparagraph\or \|\or **\or \dagger\dagger
   \or \ddagger\ddagger \else\@ctrerr\fi}}

\begin{document}
\title{Spin exchange-enabled quantum simulator for large-scale non-Abelian gauge theories}
\author{Jad C.~Halimeh${}^{*}{}^{\orcidlink{0000-0002-0659-7990}}$}
\email{jad.halimeh@physik.lmu.de}
\affiliation{Max Planck Institute of Quantum Optics, 85748 Garching, Germany}
\affiliation{Department of Physics and Arnold Sommerfeld Center for Theoretical Physics (ASC), Ludwig-Maximilians-Universit\"at M\"unchen, Theresienstra\ss e 37, D-80333 M\"unchen, Germany}
\affiliation{Munich Center for Quantum Science and Technology (MCQST), Schellingstra\ss e 4, D-80799 M\"unchen, Germany}

\author{Lukas Homeier${}^{*}{}^{\orcidlink{0000-0002-6275-6204}}$}
\email{lukas.homeier@physik.uni-muenchen.de}
\affiliation{Department of Physics and Arnold Sommerfeld Center for Theoretical Physics (ASC), Ludwig-Maximilians-Universit\"at M\"unchen, Theresienstra\ss e 37, D-80333 M\"unchen, Germany}
\affiliation{Munich Center for Quantum Science and Technology (MCQST), Schellingstra\ss e 4, D-80799 M\"unchen, Germany}
\affiliation{Department of Physics, Harvard University, Cambridge, MA 02138, USA}
\address{ITAMP, Harvard-Smithsonian Center for Astrophysics, Cambridge, MA 02138, USA}

\author{Annabelle Bohrdt${}^{\orcidlink{0000-0002-3339-5200}}$}
\affiliation{Institute of Theoretical Physics, University of Regensburg, D-93053, Germany}
\affiliation{Munich Center for Quantum Science and Technology (MCQST), Schellingstra\ss e 4, D-80799 M\"unchen, Germany}
\affiliation{Department of Physics, Harvard University, Cambridge, MA 02138, USA}
\address{ITAMP, Harvard-Smithsonian Center for Astrophysics, Cambridge, MA 02138, USA}

\author{Fabian Grusdt${}^{\orcidlink{0000-0003-3531-8089}}$}
\email{fabian.grusdt@physik.uni-muenchen.de}
\affiliation{Department of Physics and Arnold Sommerfeld Center for Theoretical Physics (ASC), Ludwig-Maximilians-Universit\"at M\"unchen, Theresienstra\ss e 37, D-80333 M\"unchen, Germany}
\affiliation{Munich Center for Quantum Science and Technology (MCQST), Schellingstra\ss e 4, D-80799 M\"unchen, Germany}

\def\thefootnote{*}\footnotetext{These authors contributed equally to this work.}

\begin{abstract}
A central requirement for the faithful implementation of large-scale lattice gauge theories (LGTs) on quantum simulators is the protection of the underlying gauge symmetry. Recent advancements in the experimental realizations of large-scale LGTs have been impressive, albeit mostly restricted to Abelian gauge groups. Guided by this requirement for gauge protection, we propose an experimentally feasible approach to implement large-scale non-Abelian $\mathrm{SU}(N)$ and $\mathrm{U}(N)$ LGTs with dynamical matter in $d+1$D, enabled by two-body spin-exchange interactions realizing local emergent gauge-symmetry stabilizer terms. We present two concrete proposals for $2+1$D $\mathrm{SU}(2)$ and $\mathrm{U}(2)$ LGTs, including dynamical bosonic matter and induced plaquette terms, that can be readily implemented in current ultracold-molecule and next-generation ultracold-atom platforms. We provide numerical benchmarks showcasing experimentally accessible dynamics, and demonstrate the stability of the underlying non-Abelian gauge invariance. We develop a method to obtain the effective gauge-invariant model featuring the relevant magnetic plaquette and minimal gauge-matter coupling terms. Our approach paves the way towards near-term realizations of large-scale non-Abelian quantum link models in analog quantum simulators.
\end{abstract}

\date{\today}
\maketitle

\textbf{\textit{Introduction.---}}Gauge theories are at the heart of the Standard Model of particle physics, comprising a well-developed framework for describing interactions between elementary particles as mediated by gauge bosons \cite{Weinberg_book,Gattringer_book,Zee_book}. They play an equally important role for describing strongly correlated quantum matter with fractionalized excitations \cite{Fradkin2013,Balents_NatureReview,Savary2016}. Paradigmatic examples of gauge theories are quantum electrodynamics with its Abelian $\mathrm{U}(1)$ gauge symmetry, the weak force with its non-Abelian $\mathrm{SU}(2)$ gauge symmetry, and quantum chromodynamics with its non-Abelian $\mathrm{SU}(3)$ gauge symmetry. These theories are currently the best descriptions of nature's three fundamental forces aside from gravity \cite{Weinberg_book}. Examples in solid state physics include strongly interacting electrons in high-temperature superconductors, which may possibly be described by Abelian $\mathrm{U}(1)$ \cite{Auerbach_book} or $\mathbb{Z}_2$ \cite{Senthil2000} gauge theories, or emergent non-Abelian $\mathrm{SU}(2)$ gauge groups \cite{Wen1996,Sachdev2019}. Nevertheless, these models remain overwhelmingly hard to solve analytically or numerically in many interesting regimes, and quantum technologies are being explored as a possible way out.

\begin{figure}[t!!]
	\centering
	\includegraphics[width=\linewidth]{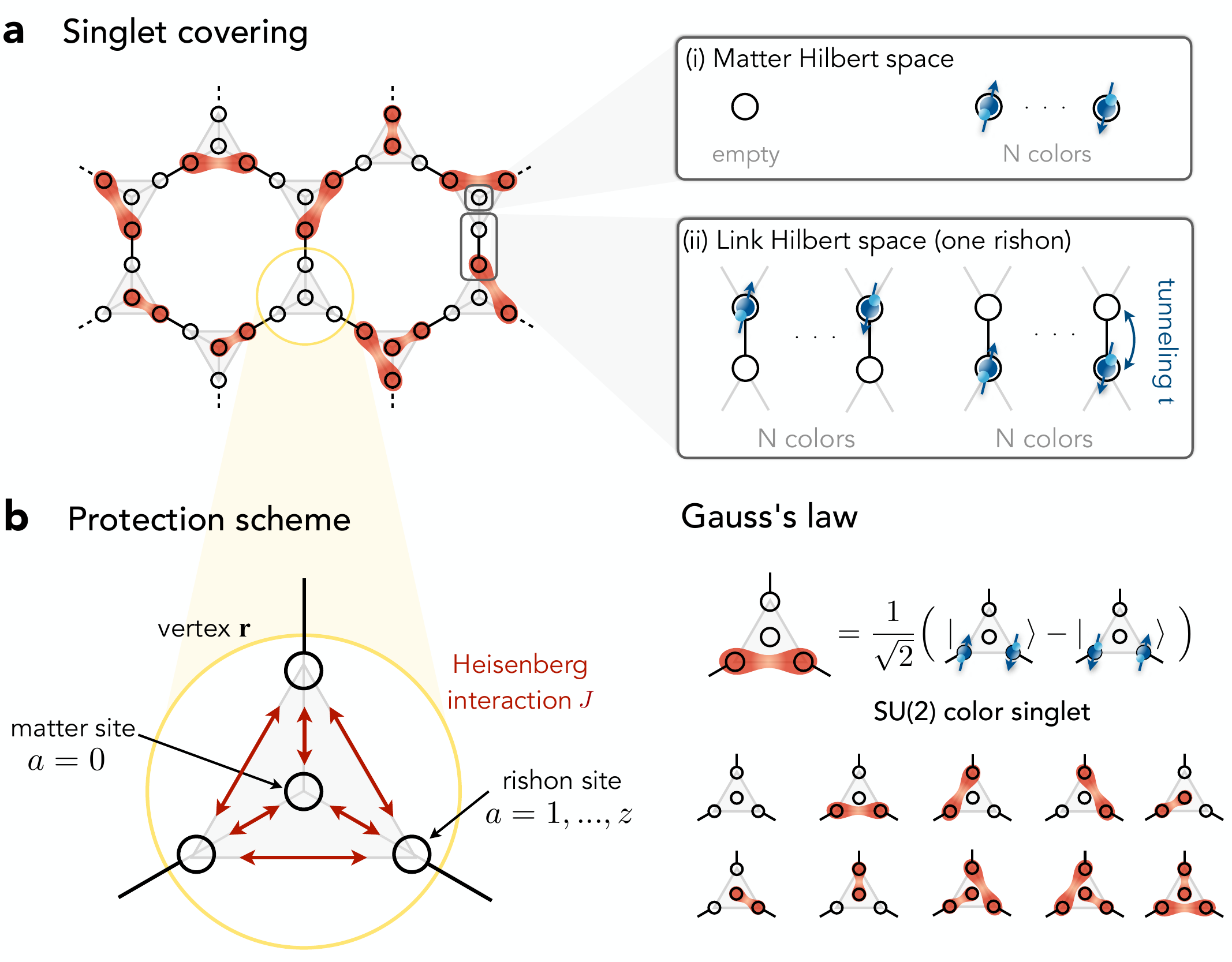}
    \caption{\textbf{$\mathrm{SU}(N)$ gauge protection.} \textbf{(a)} The rishon formulation of non-Abelian LGTs contains (i) matter on lattice sites and (ii) rishons on links connecting two sites. The links are composed of two separate rishon sites and in the simplest description we enforce to have exactly one rishon per link. The matter particles and rishons have $N$~colors; here we illustrate $N=2$. Gauge-invariant configurations in a $\mathrm{SU}(2)$~LGT have to cover the lattice with spin singlets while fulfilling the rishon number constraint. \textbf{(b)} Gauss's law enforces that the sites around a vertex have to form a total color singlet, with the physical allowed states on a vertex depicted on the right. The Gauss's law constraint can be energetically enforced by antiferromagnetic $\mathrm{SU}(N)$ gauge-invariant Heisenberg interactions between all sites adjacent to a vertex, as depicted on the left. We propose to implement the $\mathrm{SU}(2)$~LGT using ultracold atoms or molecules (blue dots). The honeycomb lattice is an appealing geometry to realize the required interaction strengths. }
\label{fig1}
\end{figure}

In recent years, a concerted effort has been undertaken to realize gauge theories on modern quantum simulators \cite{Pasquans_review,Dalmonte_review,Zohar_review,aidelsburger2021cold,Zohar_NewReview,Bauer_review,funcke2023review}. Not only can such setups provide direct experimental probes complementary to dedicated classical computations and high-energy colliders, but they also offer the exciting potential to study real-time evolution from first principles and to gain direct access to nonlocal order parameters and full counting statistics \cite{Bauer_review}.

Until now, almost all quantum-simulation experiments of gauge theories have been performed in one spatial dimension (see proposals \cite{homeier2020mathbbz2,Raychowdhury2020a,Raychowdhury2020,osborne2022largescale,Fontana2022,Homeier2022quantum,surace2023abinitio} for $2+1$D), have been restricted to a small number of degrees of freedom, or focused on Abelian gauge groups, with most non-Abelian proposals restricted to building blocks \cite{Martinez2016,Muschik2017,Bernien2017,Klco2018,Kokail2019,Schweizer2019,Goerg2019,Mil2020,Klco2020,Yang2020,Atas2021,Zhou2022,Nguyen2021,Wang2021,Mildenberger2022,Wang2022,charles2023simulating,Kadam2023,Dasgupta2022}. Partly, this is due to current technical limitations, but more importantly, extended systems are intrinsically vulnerable to gauge-breaking errors unless explicit gauge protection schemes are implemented \cite{Halimeh2020a,Halimeh_review,Homeier2022quantum}. Indeed, the only large-scale experiments so far have relied either on the complete elimination of gauge-noninvariant subspaces \cite{Martinez2016}, or on linear Stark gauge protection \cite{Yang2020,Zhou2022}. Likewise, the recent realization of a $\mathbb{Z}_2$~spin liquid in Rydberg atom arrays \cite{Semeghini2021,Wang2024_CSL} have relied on an energetically protected emergent gauge structure. 

In particular, gauge protection schemes for non-Abelian gauge theories have been few and with limited experimental feasibility \cite{Kasper2021nonabelian,Halimeh2021gauge,Mathew2022}. Experimental proposals have mostly adopted a \textit{bottom-up} approach focusing on the realization of an exact non-Abelian target gauge theory in some perturbative regime of a mapped model \cite{Banerjee2013,Tagliacozzo2013simulation,Zohar2013coldatom}, without an explicit gauge protection scheme in place. In order to further advance the quantum simulation of gauge theories, it is crucial to propose experimentally feasible realizations of large-scale non-Abelian gauge theories where gauge invariance is directly stabilized by an explicit gauge protection term that should appear naturally in experimentally relevant settings.

Here, we propose realistic implementations of non-Abelian $\mathrm{SU}(N)$ LGTs using a \textit{top-down} approach in which the focus is on realizing the gauge protection terms locally, with the gauge-invariant dynamics induced perturbatively. We demonstrate how non-Abelian gauge constraints can be energetically enforced in current and near-term quantum simulation setups based on ultracold atoms and polar molecules in optical lattices or optical tweezer arrays. Provided that these gauge protection terms define the largest energy scale in the system, the low-energy subspace is guaranteed to be described by an effective (or emergent) non-Abelian LGT. By tuning the strength or form of subdominant gauge-noninvariant terms---e.g., through tunneling or simple local spin-flip processes---the terms in the effective low-energy Hamiltonian can be conveniently controlled. Therefore, our scheme offers a realistic pathway towards large-scale implementations of non-Abelian $\mathrm{SU}(N)$ and $\mathrm{U}(N)$ lattice gauge theories, including dynamical matter and plaquette terms, and with an inherent robustness of the emergent gauge symmetry.

\textbf{\textit{Spin exchange-enabled gauge protection.---}}As a means towards experimental feasibility, \textit{quantum link formulations} of non-Abelian $\mathrm{SU}(N)$ LGTs have been proposed where the gauge-field operators are represented through fermionic degrees of freedom called \textit{rishons} \cite{Chandrasekharan1997,Wiese_review}. 
The rishon construction provides an intuitive interpretation of the gauge field: the color-charged rishons reside on the gauge-field links connecting matter sites, where each link is composed of a double well. The number $\mathcal{N}$ of rishons per link is strictly conserved, but models with different $\mathcal{N}$ can be constructed. While they all share the $\mathrm{SU}(N)$ gauge group, models with~$\mathcal{N} \rightarrow N$ are believed to more accurately resemble the continuum gauge theory.

We propose to directly implement rishon and matter sites, and energetically enforce $\mathrm{SU}(N)$ [or $\mathrm{U}(N)$] gauge invariance by a proper choice of intra-vertex interactions, where in the following we shall define a vertex. Consider a lattice with matter sites denoted by indices $\mathbf{r}$ and links $\langle\mathbf{r},\mathbf{r}'\rangle$ between nearest-neighbor sites $\mathbf{r}$ and $\mathbf{r}'$. Each link hosts two rishon sites; see Fig.~\ref{fig1}a. A vertex represented by the index $\mathbf{r}$ is then comprised of the matter site $\mathbf{r}$ and the nearest rishon sites on the links connecting to it, as illustrated in Fig.~\ref{fig1}b.

The generators of the $\mathrm{SU}(N)$ symmetry can be written in vector form as \cite{Brower1999}
\begin{align}\label{eq:Gr}
    \hat{\mathbf{G}}_\mathbf{r}=\sum_{a=0}^z\hat{\mathbf{S}}_{(\mathbf{r},a)}=\sum_{a=0}^z\sum_{\alpha,\beta=1}^N\hat{c}_{(\mathbf{r},a),\alpha}^\dagger\hat{\mathbf{T}}^{\alpha\beta}\hat{c}_{(\mathbf{r},a),\beta},
\end{align}
where $\hat{\mathbf{T}}=(\hat{T}_1,\ldots,\hat{T}_{N^2-1})$ are the $N^2-1$ elements of the $\mathrm{SU}(N)$ Lie algebra, each of which is an $N\times N$ matrix \cite{Brower1999}, $\alpha,\beta\in\{1,\ldots,N\}$ are indices representing the $N$ colors, $\hat{c}_{(\mathbf{r},a),\alpha}$ is a rishon~$(a \neq 0)$ or matter~$(a=0)$ annihilation operator on vertex~$\mathbf{r}$; the index $a\in\{0,\ldots,z\}$ indicates a matter site when $a=0$, and a ``nearest'' rishon site when $a=1,\ldots,z$ with lattice coordination number~$z$; see Fig.~\ref{fig1}b. Note that the hallmark of a non-Abelian gauge symmetry is that the $N^2-1$ components of $\hat{\mathbf{G}}_\mathbf{r}$ generally do not commute with each other. This significantly contributes to the difficulty in realizing large-scale non-Abelian gauge theories on quantum simulators using a bottom-up approach.

For a faithful gauge-theory quantum simulation, it is necessary to work in the \textit{target} or \textit{physical} gauge superselection sector $\hat{\mathbf{G}}_\mathbf{r}\ket{\psi}=0,\,\forall\mathbf{r}$. Any dynamics initialized in this sector can be restricted to remain in this sector by employing the \textit{gauge protection term} $\hat{H}_J=\frac{J}{2}\sum_\mathbf{r}\hat{\mathbf{G}}_\mathbf{r}^2$, which for $J>0$ enforces a color singlet as the ground state at each vertex \cite{Halimeh2020a,Halimeh2021gauge}, see Fig.~\ref{fig1}. The essence of our approach is to rewrite this gauge protection term in a convenient form that contains Heisenberg and Hubbard interactions, which usually naturally occur in experimental setups, thereby making it amenable for experimental implementation using ultracold molecules and atoms in an optical lattice, as we will demonstrate through concrete experimental proposals below.

In this vein, we utilize Eq.~\eqref{eq:Gr} in order to rewrite the gauge protection term as
\begin{align}\label{eq:GaugeProtection}
    \hat{H}_J=J\sum_\mathbf{r}\Bigg[\sum_{\langle a,b\rangle}\hat{\mathbf{S}}_{(\mathbf{r},a)} \cdot \hat{\mathbf{S}}_{(\mathbf{r},b)}+\frac{1}{2}\sum_a\hat{\mathbf{S}}^2_{(\mathbf{r},a)}\Bigg].
\end{align}
The first term on the right-hand side is a Heisenberg-interaction term, while the second is a two-body Hubbard term, which can be seen from the identity \cite{Haber2021}
\begin{align}\nonumber
    \hat{\mathbf{S}}_{(\mathbf{r},a)}^2=\sum_{\alpha=1}^N\bigg[&\frac{N^2-1}{2N}\hat{n}_{(\mathbf{r},a),\alpha}\\\label{eq:S2}
    &-\frac{1-\xi N}{N}\sum_{\beta;\beta > \alpha}\hat{n}_{(\mathbf{r},a),\alpha}\hat{n}_{(\mathbf{r},a),\beta}\bigg],
\end{align}
Here, $\hat{n}_{(\mathbf{r},a),\alpha}=\hat{c}_{(\mathbf{r},a),\alpha}^\dagger\hat{c}_{(\mathbf{r},a),\alpha}$ is the hard-core bosonic ($\xi=+1$) or fermionic ($\xi=-1$) number operator. In the following, we assume the matter and rishons to have the same statistics but our scheme can be easily generalized to mutual statistics.
Upon summing over all vertices, the first term of Eq.~\eqref{eq:S2} can be absorbed by the chemical potential due to the total matter plus rishon-number conservation. Taking this into account while plugging Eq.~\eqref{eq:S2} into Eq.~\eqref{eq:GaugeProtection}, we obtain the gauge protection Hamiltonian
\begin{align}\label{eq:HJ}
    \hat{H}_J&{=}\sum_{\mathbf{r},a}\bigg[J\sum_{b>a}\hat{\mathbf{S}}_{(\mathbf{r},a)}\cdot\hat{\mathbf{S}}_{(\mathbf{r},b)}{+}U\sum_{\substack{\alpha,\beta;\\\beta > \alpha}}\hat{n}_{(\mathbf{r},a),\alpha}\hat{n}_{(\mathbf{r},a),\beta}\bigg],
\end{align}
where $U=-\frac{1-\xi N}{2N}J$ now takes on the role of an on-site Hubbard interaction strength.

In this Article, we shall place the emphasis on experimentally realizing the protection term~\eqref{eq:HJ}, along with the required rishon-number conservation per link. In its general form, such a gauge protection term has been shown analytically and numerically to enable large-scale quantum simulations of \mbox{(non-)Abelian} lattice gauge theories \cite{Halimeh2020a,Halimeh2021gauge,Mathew2022}. Dynamics can then be induced perturbatively by a term $\hat{H}_t$, in part gauge-invariant and in part not, but the gauge protection term~\eqref{eq:HJ} will ensure the reliable suppression of gauge-breaking errors up to times exponential in~$J$ \cite{Halimeh2020a,Halimeh2021gauge}.

\begin{figure}[t!!]
	\centering
	\includegraphics[width=\linewidth]{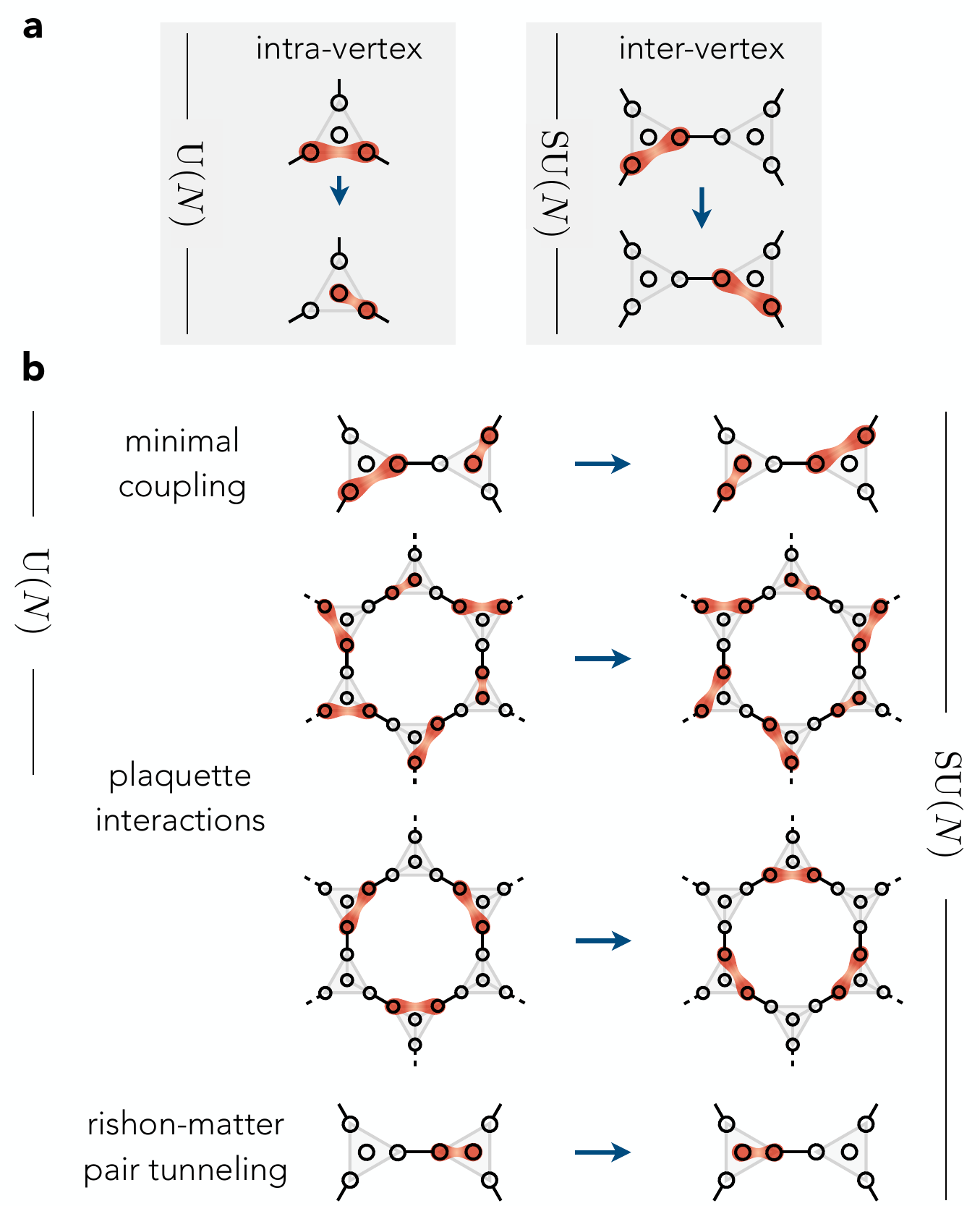}
    \caption{\textbf{Effective lattice gauge theory.} We illustrate the terms contained in the effective Hamiltonian for a one-rishon-per-link model on the honeycomb lattice, which can be realized in our proposed scheme. \textbf{(a)} $\mathrm{U}(N)$ and $\mathrm{SU}(N)$~LGTs are generated by terms that rearrange color singlets within a vertex (intra-vertex) or between neighbouring vertices (inter-vertex) involving both matter and rishon particles. The former additionally conserves a local $\mathrm{U}(1)$~symmetry. \textbf{(b)} The terms in the effective model contain the minimal gauge-matter and plaquette couplings, i.e., the kinetic and magnetic interactions of lattice gauge models. }
\label{fig2}
\end{figure}

\begin{figure*}[t!!]
	\centering
	\includegraphics[width=\linewidth]{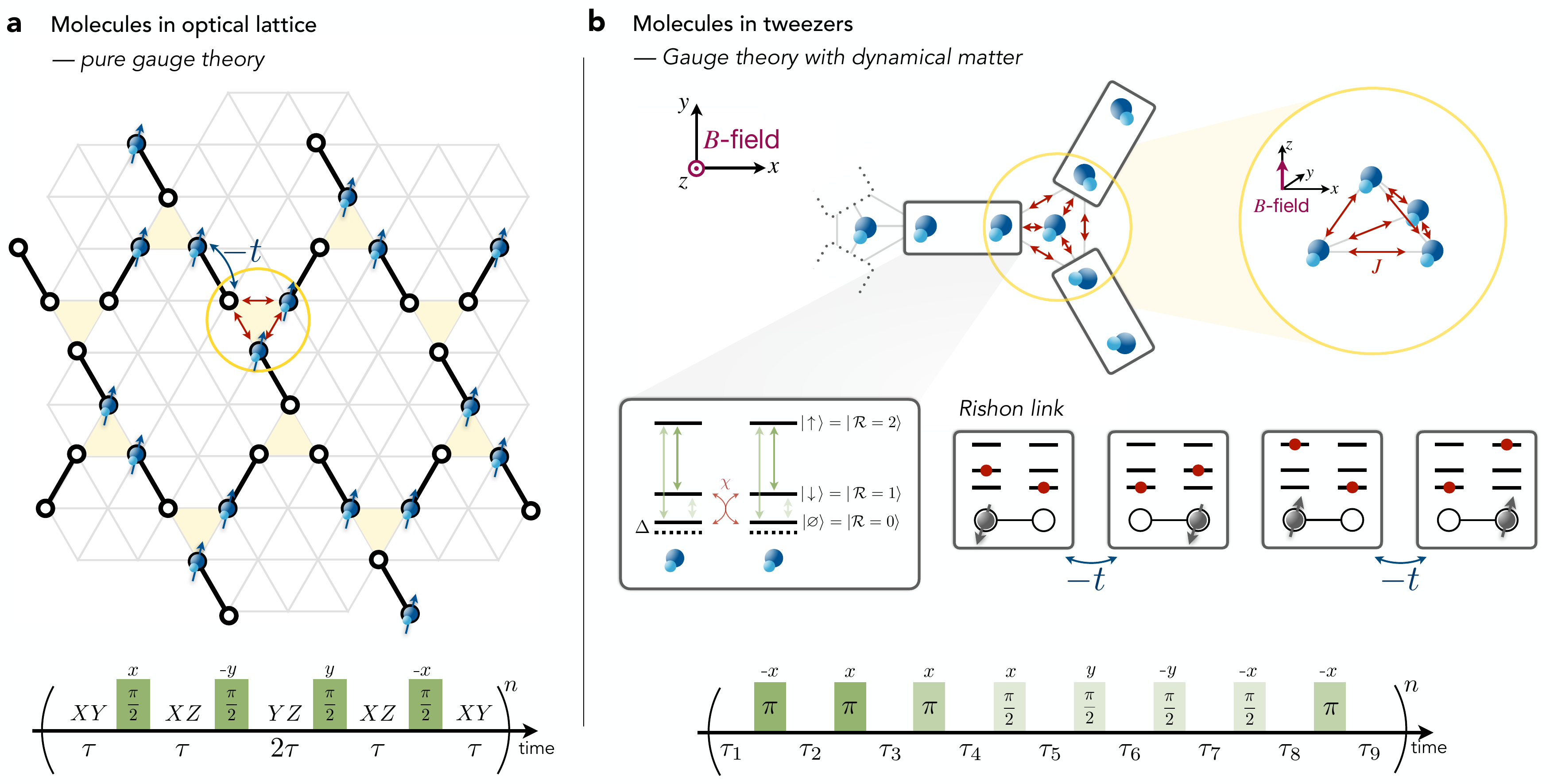}
    \caption{\textbf{Implementation with ultracold polar molecules.} \textbf{(a)} Pure $\mathrm{SU}(2)$ LGT (without dynamical matter): Ultracold molecules in optical lattices directly implement the desired microscopic Hamiltonian with gauge protection terms of strength~$J$ and weak (perturbative) tunneling~$t$. We propose to imprint a potential landscape on a triangular lattice that enables molecule tunneling only within double wells (solid black); hence the rishon number constraint is exactly fulfilled. Moreover, by using an isolated qubit subspace of internal rotational states, the dipole-dipole interaction can be used to Floquet engineer Heisenberg interactions between molecules by repeatedly applying a sequence of rotations in the qubit subspace combined with time evolution of the microscopic molecular Hamiltonian (bottom). \textbf{(b)} $\mathrm{SU}(2)$ LGT with dynamical matter: Alternatively, we propose a scheme for molecules in tweezers arrays, where the tunneling is implemented by flip-flop processes and which allows to include dynamical matter. Each link is built from two molecules and the molecule on the matter site is located in a second plane such that the interaction strength around molecules adjacent to a vertex have equal strength~$J$. We encode the states~$\ket{\varnothing}$, $\ket{\downarrow}$, and $\ket{\uparrow}$ in each molecule's internal rotational states~$\ket{\mathcal{R}=0}$, $\ket{\mathcal{R}=1}$, and $\ket{\mathcal{R}=2}$, respectively. By applying a Floquet sequence which rotates within the molecular rotational states, both rishon tunneling as well as the Heisenberg interactions can be realized efficiently. To avoid tunneling between different rishon links or between matter and rishon sites, we require local detunings $\Delta$ for the $\ket{\mathcal{R}=0}$ states by local AC stark shifts (see Appendix~\ref{sec:bosonic}).}
\label{fig3}
\end{figure*}

\textbf{\textit{Effective lattice gauge theory.---}}The effective (gauge-invariant) Hamiltonian emerges from the microscopic Hamiltonian $\hat{H}=\hat{H}_J+\hat{H}_t$, which we will further discuss below in experimentally relevant settings. 
To recap, our method is based on a separation of energy scales with (i) strong protection terms~\eqref{eq:GaugeProtection} of strength~$J$ and (ii) weak tunneling perturbation~$t$, allowing us to treat the latter perturbatively and to derive the effective gauge-invariant model (see Appendix~\ref{sec:effective}).
The perturbation describes the tunneling of rishons within links as well as tunneling of matter between vertices.

For sufficiently weak perturbations, $\lvert t\rvert \ll \lvert J\rvert$, the system approaches a controlled-violation regime, where the occupation of gauge-noninvariant sectors is strongly suppressed. Hence, the low-energy effective theory is inherently gauge-invariant, which we confirm by numerical simulations of small systems (see Appendix~\ref{sec:effective}).
This allows us to determine the dominant terms of the effective Hamiltonian, which are gauge-invariant and conserve the rishon number constraint.
In our model, we identify two types of processes:

Firstly, the simultaneous rearrangement of color singlets within a vertex (intra-vertex) while maintaining the rishon and matter number constraint is manifestly gauge-invariant, see Fig.~\ref{fig2}a.
Since the total number of particles around a given vertex is maintained, the terms generated by intra-vertex processes host an additional local $\mathrm{U}(1)$~symmetry, which yields an effective $\mathrm{U}(N)$~LGT~\cite{Brower1999}.

Secondly, our method induces terms that are $\mathrm{SU}(N)$ gauge-invariant but break the local $\mathrm{U}(1)$~symmetry.
Thus, the local symmetry structure is reduced to $\mathrm{SU}(N)$.
The relevant $\mathrm{SU}(N)$ gauge-invariant terms involve inter-vertex tunneling of matter and rishons, in which whole color singlets tunnel between vertices; see Fig.~\ref{fig2}a.

We emphasize that our protection-based approach offers the flexibility to realize both $\mathrm{U}(N)$ and $\mathrm{SU}(N)$~LGTs.
By introducing disordered protection strength, i.e., vertex-depended protection~$J \rightarrow J_\mathbf{r}$, the inter-vertex processes are suppressed. 
In multi-rishon models, the $\mathrm{SU}(N)$ gauge-invariant model further contains terms that transport glueballs, i.e., bound states of rishon color singlets~\cite{Wiese_review}.

\textbf{\textit{One-rishon models: ultracold molecules.---}}In order to demonstrate our approach, we shall first focus on the case of large-scale $2+1$D $\mathrm{SU}(2)$ LGTs. For the purpose of most realistic and immediate experimental relevance, we further employ a hard-core constraint on the matter and rishon sites, while fixing the rishon number per link to unity, $\mathcal{N}=1$. 
This allows us to drop the second term in Eq.~\eqref{eq:HJ}, and only focus on proposing an experimentally feasible scheme for the realization of the Heisenberg term. The latter involves equal-strength magnetic interactions at magnitude $J$ around a vertex, and a rishon per link that can tunnel between the two sites on any given rishon link. This tunneling, along with a similar term for the matter, does not commute with $\hat{H}_J$ and will take the role of the gauge-breaking perturbation with strength $ \lvert t\rvert \ll \lvert J\rvert$.

For models with exactly one rishon per site, the statistics of the rishon has no effect as long as we assume that rishons cannot tunnel between different rishon links.
In this case, we can replace the fermionic rishons by bosons.

Ultracold polar molecules are a new powerful tool~\cite{Yan2013, Wall2015_molecules,Luo2020,Gregory2021, Schindewolf2022,Holland2022,Bigagli2023} to implement and control $\mathrm{SU}(2)$-invariant Heisenberg interactions by Floquet-driving the intrinsic dipole-dipole interactions \cite{Christakis2023}. Recently, also coherent tunneling of ultracold molecules in optical lattices in the presence of dipolar spin interactions has been achieved~\cite{Carroll2024}. In the following, we suggest two different schemes below: with and without physical tunneling of molecules. Furthermore, we note that molecules currently face problems to reach high fillings of tweezers/optical lattices, and deterministic filling is not (yet) possible, but we expect that such technical challenges can soon be overcome~\cite{Picard2024}.

\textit{Optical lattice: Tunneling scheme.---}Let us first consider an $\mathrm{SU}(2)$ LGT \textit{without} dynamical matter, and with one rishon per link.
The use of ultracold molecules in optical lattices has the advantage that the local Hilbert space structure and hence the rishon-number constraint can be exactly fulfilled by simply suppressing tunneling outside the links for all experimentally relevant times.
To this end, we propose to implement an optical landscape with double-well potentials on links between vertices either by a superlattice or by removing lattice sites from an optical lattice using a digital mirror device (``cookie-cutting method''), see Fig.~\ref{fig3}a. Each double well should be loaded with exactly one molecule, and we restrict the molecular internal states to a qubit subspace to encode the two colors~$\downarrow$ and~$\uparrow$.

Thus, the microscopic Hamiltonian we suggest to implement is given by~$\hat{H} = \hat{H}_{J} + \hat{H}_{t}$ with the (weak) rishon-tunneling perturbation
\begin{align} \label{eq:tunneling}
    \hat{H}_{t} = -t \sum_{\langle \mathbf{r}, \mathbf{r}' \rangle } \sum_{\alpha=\downarrow,\uparrow}\Big[\hat{c}^\dagger_{(\mathbf{r},a),\alpha}\hat{c}_{(\mathbf{r}',b),\alpha} + \mathrm{H.c.}\Big],
\end{align}
where the indices~$a, b$ defining the rishon link are uniquely determined by the link $\langle \mathbf{r}, \mathbf{r}'\rangle$. The geometry has to be chosen such that the sites around a vertex are pairwise equally distanced.
A chain in $1+1$D trivially fulfills this constraint;
in $2+1$D, this can be achieved in, e.g., a honeycomb lattice as shown in Fig.~\ref{fig1}a, i.e., the rishon links form a kagome lattice. Other lattices would be feasible by either exploiting out-of-plane geometries or anisotropic interactions.

In our proposed setup in $2+1$D, the Gauss's law constraint yields a covering of singlets on the kagome lattice, which is a highly degenerate ground-state manifold.
In the ground-state, this particular model can be re-written in terms of a quantum dimer model on a non-bipartite lattice~\cite{Banerjee2018} hosting a $\mathbb{Z}_2$~spin liquid phase in the deconfined regime~\cite{Verresen2021, Semeghini2021}.
We emphasize, however, that our model features distinctly different vertex excitations associated with the stabilizer term~$\hat{H}_J$ if we allow individual gauge-breaking excitations that carry well-defined $\mathrm{SU}(2)$~charges.

The gauge protection~\eqref{eq:HJ} can be implemented by using strong nearest-neighbor dipole-dipole interactions as follows: molecules located around the same vertex interact with resonant dipole-dipole interactions of strength~$\chi$ between two rotational states of molecules giving rise to XY~spin interactions.
In order to engineer the $\mathrm{SU}(2)$ gauge-invariant Heisenberg interactions, which energetically enforce the low-energy manifold to fulfill the Gauss's law constraint, see Fig.~\ref{fig1}b, we propose to use Floquet-driving of the system~\cite{Geier2021, Scholl2022, Christakis2023}:
Consecutive $\pi/2$-rotations around the Pauli $(x, -y, y, x)$-direction yield an effective isotropic Heisenberg Hamiltonian in the limit of high frequencies~$1/T \gg \chi $, where~$T$ is the Floquet cycle time, see Fig.~\ref{fig3}a.

To induce dynamics within the gauge-invariant subspace, we propose to weakly perturb the system with gauge-breaking terms. 
Due to the energetic constraints, the system evolves under an $\mathrm{SU}(2)$ gauge-invariant Hamiltonian and undesired gauge sectors remain only virtually occupied \cite{Halimeh2021gauge}.
In our scheme, the tunneling~$\lvert t\rvert \ll \lvert J\rvert$ between neighboring rishon sites is a sufficient perturbation to induce the desired dynamics, which we numerically benchmark below.

We remark that the dipolar interactions decay with the cube of the distance between two molecules giving rise to long-range interactions.
Thus, beyond nearest-neighbor, interactions can be treated as another gauge-breaking perturbation with a strength much smaller than~$\lvert J\rvert$; in the presence of Floquet driving the perturbation is long-range Heisenberg-type.
In our proposed scheme with built-in gauge protection, the extra perturbation yields an additional source to induce dynamics within the gauge-invariant subspace.

Dynamical matter can be included in principle by introducing an additional lattice, e.g., in a bilayer geometry, on which matter molecules of the same species interact with rishon molecules around a vertex with Heisenberg interactions at strength $J$.
Tunneling~$\lvert t\rvert \ll \lvert J\rvert$ allows the molecules to move between matter sites.
While the tunneling scheme is conceptually the simplest model, realizing it with dynamical matter in current setups is challenging.

\textit{Tweezer arrays: Bosonic excitation tunneling scheme.---}We propose a second scheme that implements tunneling through flip-flop processes of bosonic (spin-flip) excitations by including a third rotational molecular state to encode an empty site. Here, we assume a fully filled lattice with one molecule on each matter/rishon site.
To engineer both tunneling and magnetic interactions, we adapt a Floquet driving scheme that is based on resonant dipole-dipole interactions combined with rotations within the subspace of three internal states, which has been recently proposed in Ref.~\cite{Homeier2023} (see Appendix~\ref{sec:bosonic}).

For concreteness, we also assume that the matter sites form a honeycomb lattice with adjacent rishon links. Since we do not require physical tunneling between any sites on the lattice, the geometry can be realized with optical tweezers.
The link building block, i.e., two sites occupied by one bosonic rishon, is constructed from two molecules:
For the left and right molecule, we identify the internal rotational molecular states~$\ket{\mathcal{R}}$ with: an empty site~$\ket{\varnothing}$ ($\mathcal{R}=0$), a $\downarrow$-rishon ($\mathcal{R}=1$), and a $\uparrow$-rishon ($\mathcal{R}=2$).
We will describe below how the effective dynamics can be restricted to a subspace that fulfills the rishon number constraint; see Fig.~\ref{fig2}b.
Additionally, we place molecules on matter sites and identify the internal states with bosonic matter fields as in the rishon case.
To obtain equal-strength interactions around a vertex on, e.g., the honeycomb lattice, we propose to elevate the molecules on matter sites adequately out-of-plane~\cite{Barredo2018}.

Using this mapping, the Heisenberg interaction term~$J$ required for gauge protection is engineered with the same Floquet driving sequence as proposed for the optical lattice implementation.
Additionally, the tunneling within a rishon link corresponds to an exchange processes $\ket{\mathcal{R}=0, \mathcal{R}=1} \leftrightarrow \ket{\mathcal{R}=1, \mathcal{R}=0}$ ($\ket{\mathcal{R}=0, \mathcal{R}=2} \leftrightarrow \ket{\mathcal{R}=2, \mathcal{R}=0}$) for tunneling of a $\downarrow$-rishon ($\uparrow$-rishon).
The $\downarrow$-rishon tunneling is conveniently implemented by the resonant dipole-dipole interaction. In order to obtain the same tunneling strength for the $\uparrow$-rishon, an extra rotation within the Floquet sequence has to be included. We refer to the Appendix~\ref{sec:bosonic}, for details.

\begin{figure}[t!!]
	\centering
	\includegraphics[width=\linewidth]{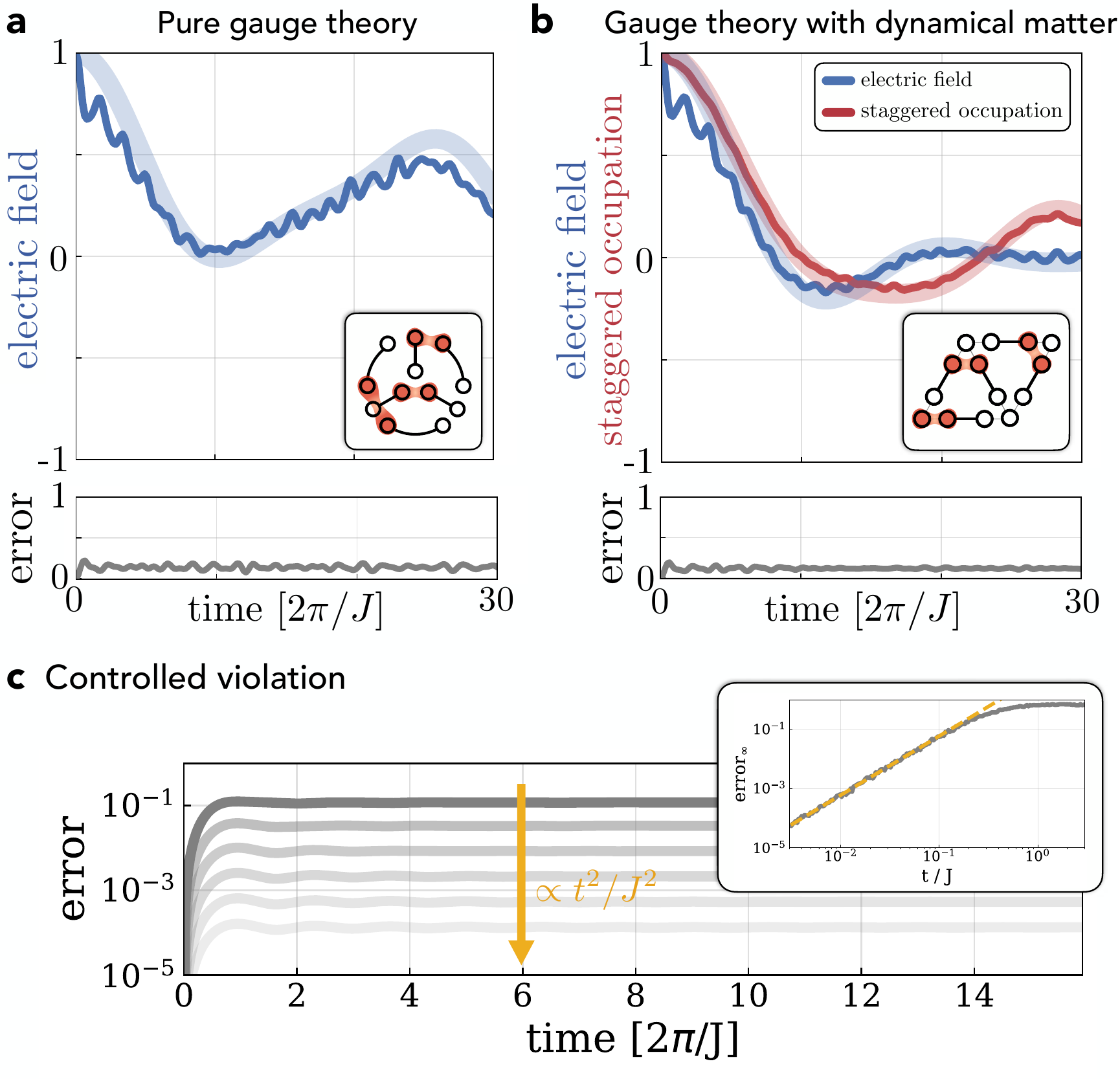}
    \caption{\textbf{Exact diagonalization study.} \textbf{(a)} Quench dynamics of the electric field in the $\mathrm{SU}(2)$ pure LGT without dynamical matter on a Mercedes star geometry for $t/J=0.075$. The initial state is illustrated in the inset. Throughout the entire dynamics, the gauge violation is always below $22\%$ of the infinite-temperature violation (which corresponds to an error of $1$ in our plot), settling into a stable plateau $\propto t^2/J^2$ (see Fig.~\ref{figGaugeViolation} in Appendix~\ref{sec:effective}). \textbf{(b)} Quench dynamics of the electric field in the $\mathrm{SU}(2)$ LGT with dynamical matter on a double-triangle geometry with four matter sites at the corners for $t=4t_m=0.15J$. The initial state is illustrated in the inset. Throughout the entire dynamics, the gauge violation is always below $20\%$ of the infinite-temperature violation.  Shaded lines in (a) and (b) correspond to dynamics of the corresponding gauge-invariant effective Hamiltonian, which agree well with our exact prediction for the microscopic model. In both cases, the microscopic model exhibits rich gauge-invariant dynamics in a controlled-violation regime. In \textbf{(c)}, we plot the gauge violation for $t=4t_m=0.15\times2^{-k}J$ with $k=0,1,\ldots,5$ (dark to bright gray) for the setup discussed in \textbf{(b)}, which settles into a stable plateau $\propto t^2/J^2$ (see inset).}
\label{fig4}
\end{figure}

To suppress hopping of rishon excitations between \textit{different} links or onto matter sites, local AC stark shifts~$\lvert\Delta_j\rvert \gg 1/T$ can be applied to the $\ket{\mathcal{R}=0}$ states, such that resonant tunneling~$t$ only occurs if~$\Delta_i = \Delta_j$, where $i$ and $j$ are indices distinguishing between the different links or matter sites.
In the effective Floquet model, the coupling ratio~$t/J$ can be tuned by adjusting the distance between molecules around a vertex and between the two rishon sites.
Hence, the tunneling rate between matter sites $t_m$ and between rishon sites $t$ is determined by the geometry (see Appendix~\ref{sec:bosonic}). Concretely, the microscopic Hamiltonian is now $\hat{H}=\hat{H}_J+\hat{H}_t+\hat{H}_{t_m}$, where
\begin{align} \label{eq:tunneling_matter}
    \hat{H}_{t_m} = -t_m\sum_{\langle \mathbf{r}, \mathbf{r}' \rangle} \sum_{\alpha=\downarrow,\uparrow}\Big[\hat{c}^\dagger_{(\mathbf{r},0),\alpha}\hat{c}_{(\mathbf{r}',0),\alpha} + \mathrm{H.c.}\Big].
\end{align}

\textit{Numerical benchmarks.---}We now validate our scheme by considering the two aforementioned $\mathrm{SU}(2)$ LGTs (the one with dynamical matter and the one without) on geometries amenable for exact diagonalization (ED). The quench dynamics of the electric field, defined as the local staggered rishon occupation on links, is shown in Fig.~\ref{fig4} for the $\mathrm{SU}(2)$ LGT without matter on a \textit{Mercedes star} geometry with $t=0.075J$ and for the $\mathrm{SU}(2)$ LGT with dynamical matter on a double-triangle geometry with four matter sites at the corners and with $t=4t_m=0.15J$. In both cases, the initial state is shown as an inset. For the case with dynamical matter in Fig.~\ref{fig4}b, we additionally show the difference in matter occupation between matter sites on the long diagonal and those on the short diagonal, with only one matter site on the long diagonal initially occupied. For all local observables, we find rich fast dynamics within experimentally friendly evolution times. Crucially, the parameters used for the tunneling are in a controlled-violation regime, guaranteeing faithful gauge-theory dynamics up to times exponential in $J$ \cite{Halimeh2020a}. Indeed, for both models we find that the gauge violation, defined as $\sum_\mathbf{r}\langle\hat{\mathbf{G}}_\mathbf{r}^2\rangle$ normalized by its infinite-temperature value, to be suppressed into a plateau $\propto t^2/J^2$ for all investigated late times (see also Fig.~\ref{figGaugeViolation} in Appendix~\ref{sec:effective}), and is always below $22\%$ for the parameters employed in Fig.~\ref{fig4}.

The dynamics from the microscopic model $\hat{H}$ for the parameters employed in Fig.~\ref{fig4} is faithfully reproduced by an effective gauge-invariant Hamiltonian $\hat{H}_\text{eff}$. The latter is extracted numerically by projecting the time-evolution unitary propagated by $\hat{H}$, for some evolution interval $\delta\tau$, onto the gauge-invariant color-singlet subspace, $\hat{\mathcal{P}}_Ge^{-i\hat{H}\delta\tau}\hat{\mathcal{P}}_G\approx e^{-i\hat{H}_\text{eff}\delta\tau}$, where $\hat{\mathcal{P}}_G$ is the corresponding projector. An optimal value for $\delta\tau$ is then found in the range $1/J\ll\delta\tau\ll J/t^2$ at which the dynamics of both models show very good quantitative agreement, which allows constructing the matrix elements of $\hat{H}_\text{eff}$. We find that $\hat{H}_\text{eff}$ contains minimal coupling and plaquette terms (see Appendix~\ref{sec:effective} for details), which is a very significant result given that the implementation of plaquette terms in proposals of large-scale quantum simulators of gauge theories face severe experimental limitations \cite{osborne2022largescale,surace2023abinitio}.

\begin{figure}[t!!]
	\centering
	\includegraphics[width=\linewidth]{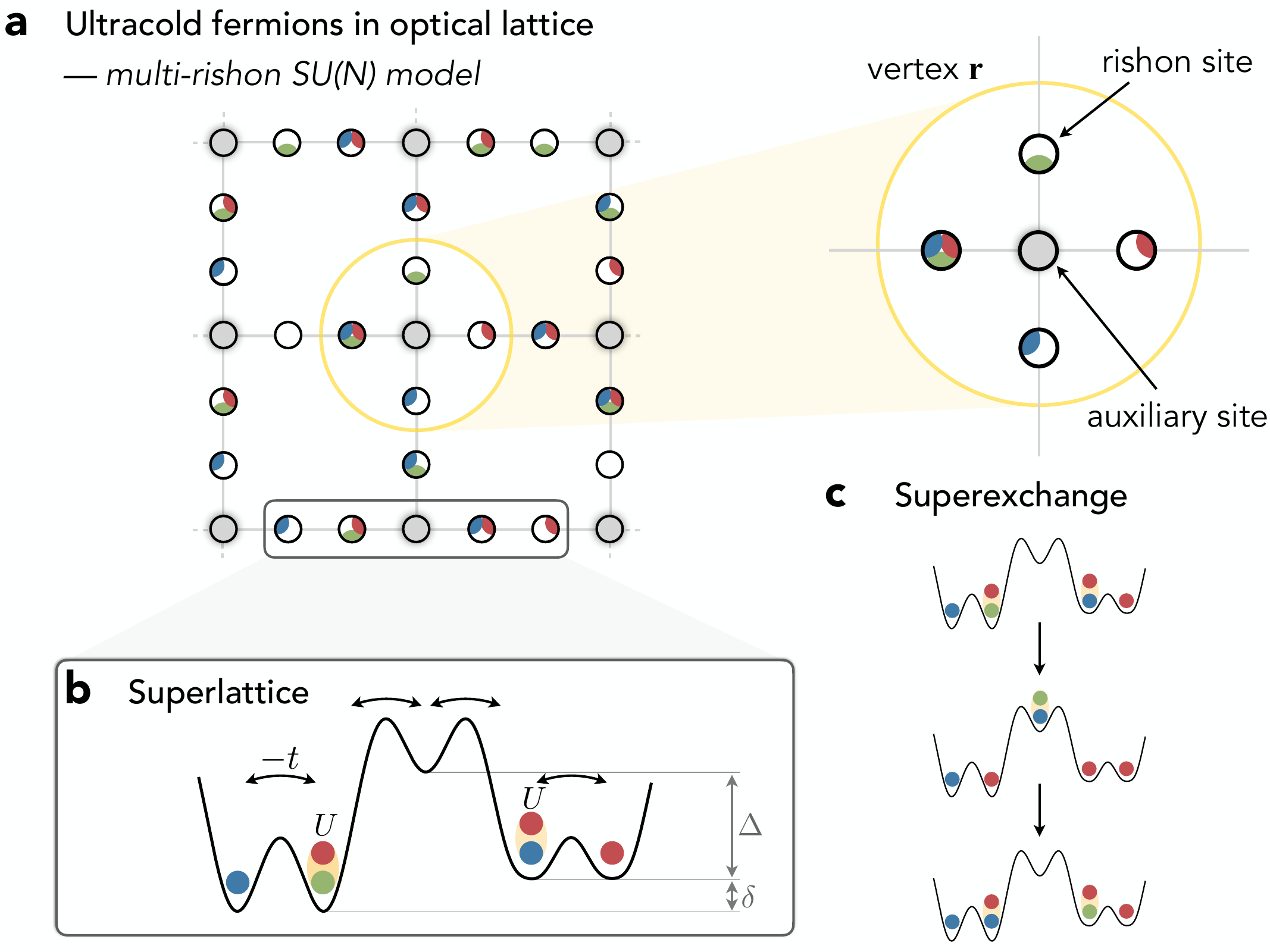}
    \caption{\textbf{Multi-rishon model.} \textbf{(a)} An arbitrary number of rishons encoded in fermionic alkaline-earth atoms are placed on rishon links formed by double-well potentials; here we illustrate an $\mathrm{SU}(3)$ case with colors red, blue and green, and three rishons per link. \textbf{(b)} The atoms can tunnel between neighboring sites with amplitude~$t$ and interact with on-site attractive Hubbard interactions~$U<0$. An optical superlattice prevents direct tunneling of rishons between different links by applying a potential offset~$\lvert \delta\rvert \gg \lvert t\rvert, \lvert U\rvert$. Each rishon link is connected to an auxiliary site with large potential offset~$\lvert \Delta\rvert \gg \lvert t\rvert, \lvert U\rvert$. \textbf{(c)} Fourth-order processes, in which atoms of neighboring links virtually occupy the auxiliary site and interact with Hubbard interactions, yield the desired equal-strength antiferromagnetic interaction between all sites connected to a vertex~$\mathbf{r}$. Together with the on-site Hubbard interactions, the effective model includes all required building blocks to gauge-stabilize a multi-rishon-per-link $\mathrm{SU}(N)$~LGT, in which the dynamics is perturbatively induced by tunneling~$t$ within rishon links.  }
\label{fig5}
\end{figure}

\textbf{\textit{Multi-rishon-per-link models: $\mathrm{SU}(N)$ with alkaline-earth atoms.---}}The long-sought goal is the realization of $\mathrm{SU}(N)$ LGTs beyond one rishon per link and with fermionic matter.
Now, we propose a possible scheme to implement the required gauge protection, which in principle enables the large-scale simulation of lattice QCD. 
In the multi-rishon regime, we cannot neglect the second term in Eq.~\eqref{eq:HJ} nor the fermionic statistics on rishon links.
Hence, gauge protection requires a combination of on-site attractive Hubbard interactions~$U<0$ and antiferromagnetic interactions~$J>0$ around all sites adjacent to a vertex.

(i) We envision a scheme based on an $\mathrm{SU}(N)$ Fermi-Hubbard model of alkaline-earth atoms~\cite{Hofrichter2016, Taie2022} in an optical superlattice combined with tweezer arrays.
As illustrated in Fig.~\ref{fig5}a, the rishon links are built from double-well potentials attached to an auxiliary central site with potential offset~$\Delta$.
The potential landscape with additional offsets~$\delta$ between rishon links shown in Fig.~\ref{fig5}b, ensures the conservation of rishon number per link.

A fourth-order superexchange mechanism, where atoms can virtually occupy the auxiliary site, yields the desired $\mathrm{SU}(N)$-invariant antiferromagnetic interactions around a vertex mediated by Hubbard interactions, see Fig.~\ref{fig5}c.
Moreover, attractive Hubbard interactions~$U<0$ accommodate for the required on-site term in Eq.~\eqref{eq:HJ} and their strength is tunable via a Feshbach resonance.
We note that the detuning~$\Delta$ between the rishon and auxiliary sites allows to tune the sign of the superexchange~$J$ to obtain antiferromagnetic couplings (see Appendix~\ref{sec:multi}). In Appendix~\ref{sec:multi-numerics}, we numerically simulate a minimal model composed of two links and two fermionic $\mathrm{SU}(2)$ rishons per link demonstrating the efficiency of our predicted fourth-order superexchange gauge protection scheme, and the applicability of the top-down approach for multi-rishon models.

Again, gauge invariant dynamics is induced by weak perturbations given by the tunneling between rishon and matter sites in the optical lattice.
We emphasize that the proposed scheme is independent of the coordination number or geometry since the interactions are mediated through the on-site Hubbard interaction on the auxiliary site in contrast to the long-range dipole-dipole interactions described above.
Moreover, $\mathrm{SU}(N)$ color charges can be conceptually easily included in this scheme by introducing a second layer, where atoms are located on matter sites connected to the auxiliary site of its corresponding vertex.

(ii) Another approach to realize multi-rishon models are hybrid analog-digital fermionic quantum processors~\cite{fermiqp,Hartke2022,GonzalezCuadra2023}, which potentially allow for much stronger interaction scales than the superexchange-based scheme~\cite{Nemirovsky2021}.
In these platforms, fermionic atoms are trapped in optical tweezers that are spatially rearranged throughout the quantum simulation.
A combination of interaction and tunneling gates can then realize the desired Hamiltonian we propose here with protection and perturbation terms, respectively.
The arbitrary connectivity enables to include fermionic, dynamical matter or implement models beyond $2+1$D.

\textbf{\textit{Summary and outlook.---}}We have presented a \textit{top-down} approach for realizing large-scale $d+1$D $\mathrm{U}(N)$ and $\mathrm{SU}(N)$ lattice gauge theories in setups of ultracold polar molecules and alkaline-earth atoms. Our method focuses on utilizing the naturally arising Heisenberg and Hubbard interactions in such setups in order to enforce the local color singlet subspaces constituting Gauss's law at the vertices. The gauge-theory dynamics is then induced perturbatively through a gauge-noninvariant tunneling term, which leads to an effective lattice gauge theory with minimal coupling and plaquette terms. This effective gauge-invariant model is stabilized with its gauge violation suppressed up to all experimentally and numerically relevant times.

To showcase the utility of our scheme, we have proposed concrete $2+1$D $\mathrm{SU}(2)$ LGTs with and without dynamical matter and with a one-rishon-per-link constraint that can be readily realized in optical lattices or in setups of ultracold molecules with tweezer arrays. We performed numerical benchmarks establishing rich gauge-invariant dynamics over experimentally relevant timescales, with a controlled and well-suppressed gauge violation over all accessible evolution times. Such proposals are viable for implementation with current state-of-the-art quantum-simulator technology.

We further outlined a proposal for the realization of multi-rishon-per-link $\mathrm{SU}(N)$ LGTs in platforms of alkaline-earth atoms and also on hybrid analog-digital fermionic processors, which may facilitate realizations beyond two spatial dimensions by reaching higher protection-energy scales and richer dynamics at shorter timescales.

Our \textit{top-down} approach of enforcing Gauss's law locally and perturbatively inducing effective gauge-invariant dynamics, which we explicitly extract, opens the door towards large-scale quantum simulators of non-Abelian LGTs in a realistic and controlled fashion.

\textbf{\textit{Acknowledgments.---}}We are grateful for inspiring discussions with M.~Aidelsburger, L.~Cheuck, P.~Hauke, T.~A.~Hilker, A.~Kaufman, K.-K.~Ni, and A.~Park.
L.H.~acknowledges support from the Studienstiftung des deutschen Volkes.
This research is funded by the European Research Council (ERC) under the European Union’s Horizon 2020 research and innovation programm (Grant Agreement no 948141) — ERC Starting Grant \mbox{SimUcQuam}, by the Deutsche Forschungsgemeinschaft (DFG, German Research Foundation) under Germany's Excellence Strategy -- EXC-2111 -- 390814868 and via Research Unit FOR 2414 under project number 277974659, by the NSF through a grant for the Institute for Theoretical Atomic, Molecular, and Optical Physics at Harvard University and the Smithsonian Astrophysical Observatory. This work was supported by the QuantERA grant DYNAMITE, by the Deutsche Forschungsgemeinschaft (DFG, German Research Foundation) under project number 499183856, and within the QuantERA II Programme that has received funding from the European Union’s Horizon 2020 research and innovation programme under Grand Agreement No 101017733.
This research was supported by the Munich Institute for Astro-, Particle and BioPhysics (MIAPbP) which is funded by the Deutsche Forschungsgemeinschaft (DFG, German Research Foundation) under Germany's Excellence Strategy – EXC-2094 – 390783311.

\textbf{\textit{Author contributions.---}}J.C.H.~conceived the idea of dynamical gauge protection. J.C.H., L.H.,~and F.G.~adapted the idea to experimentally relevant non-Abelian systems. L.H.~devised the experimental proposal for alkaline-earth atoms, with help from F.G. L.H.~devised the experimental proposal for ultracold molecules, with help from A.B. J.C.H., L.H. and F.G. performed the numerical simulations. L.H.~performed all analytic derivations. J.C.H.~and L.H.~wrote the bulk of the manuscript, with all authors contributing to its preparation.

\appendix

\section{$\mathrm{SU}(N)$~gauge protection}\label{sec:gaugeprotection_appendix}

The gauge protection term~\eqref{eq:GaugeProtection} can be derived by manipulating the $\mathrm{SU}(N)$~symmetry generators in the rishon formulation.
The color singlet constraint can be energetically enforced by the Hamiltonian
\begin{align}
    \hat{H}_J = \frac{J}{2} \sum_{\mathbf{r}} \hat{\mathbf{G}}_\mathbf{r}^2
\end{align}
with~$J>0$.
We can re-write each vertex~$\mathbf{r}$ as
\begin{align}
    \hat{\mathbf{G}}_\mathbf{r}^2 = 2\sum_{\langle a,b \rangle} \hat{\mathbf{S}}_{(\mathbf{r},a)}\cdot\hat{\mathbf{S}}_{(\mathbf{r},b)} + \sum_{a} \hat{\mathbf{S}}^2_{(\mathbf{r},a)}
\end{align}
The first term describes the all-to-all Heisenberg interactions.
The second term enforces on-site spin singlets.
We express the on-site term via two-body Hubbard interactions for $\mathrm{SU}(N)$~spins (we suppress the site index here):
\begin{align}
    \begin{split}
        \hat{\mathbf{S}}^2 &= \sum_{\alpha, \beta} \sum_{\gamma, \delta} \hat{c}^\dagger_{\alpha}\hat{c}_{\beta}\hat{c}^\dagger_{\gamma}\hat{c}_{\delta} \left( \vec{T}_{\alpha,\beta} \cdot \vec{T}_{\gamma, \delta} \right) \\
        &= \sum_{\alpha, \beta} \sum_{\gamma, \delta} \hat{c}^\dagger_{\alpha}\hat{c}_{\beta}\hat{c}^\dagger_{\gamma}\hat{c}_{\delta} \frac{1}{2} \left( \delta_{\alpha,\delta}\delta_{\beta,\gamma} - \frac{1}{N}\delta_{\alpha\beta}\delta_{\gamma,\delta} \right) \\
        &= \frac{1}{2} \sum_{\alpha, \beta} \left( \hat{c}^\dagger_{\alpha}\hat{c}_{\beta} \hat{c}^\dagger_{\beta} \hat{c}_{\alpha} - \frac{1}{N} \hat{c}^\dagger_{\alpha}\hat{c}_{\alpha} \hat{c}^\dagger_{\beta}\hat{c}_{\beta}   \right) \\
        &= \frac{1}{2} \sum_{\alpha, \beta} \left[ \hat{c}^\dagger_{\alpha}\left(1+\xi \hat{c}^\dagger_{\beta}\hat{c}_{\beta}\right) \hat{c}_{\alpha} - \frac{1}{N} \hat{n}_{\alpha}\hat{n}_{\beta}   \right] \\
        &= \frac{N}{2} \sum_{\alpha} \hat{n}_{\alpha}+\frac{\xi}{2}\sum_{\alpha, \beta} \hat{c}^\dagger_{\alpha}\hat{c}^\dagger_{\beta}\hat{c}_{\beta} \hat{c}_{\alpha} - \frac{1}{2N}\sum_{\alpha, \beta}\hat{n}_{\alpha}\hat{n}_{\beta}\\
        &= \frac{N^2}{2N} \sum_{\alpha} \hat{n}_{\alpha}+\frac{\xi N}{2N}\sum_{\substack{\alpha,\beta;\\\beta > \alpha}} \hat{n}_{\alpha}\hat{n}_{\beta} - \frac{1}{2N}\sum_{\alpha, \beta}\hat{n}_{\alpha}\hat{n}_{\beta}\\
        &= \frac{N^2-1}{2N} \hat{n} - \frac{1 - \xi N}{N}\sum_{\substack{\alpha,\beta;\\\beta > \alpha}} \hat{n}_\alpha \hat{n}_\beta
    \end{split}
    \label{eq:totalS_to_Hubbard}
\end{align}
with $\hat{n} = \sum_\alpha \hat{n}_\alpha$.
Further, we have used the hard-core constraint of fermions~$(\xi=-1)$ or hard-core bosons~$(\xi=+1)$, i.e., $\hat{n}_\alpha^2 = \hat{n}_\alpha$.
We find that the on-site spin singlets are enforced by an (unimportant) chemical potential and a Hubbard interaction with~$U=-\frac{1 - \xi N}{2N}J$; see Eq.~\eqref{eq:HJ}.
The sign of the interaction for fermions with~$N>1$ is negative and thus attractive.

\section{Effective gauge theory}\label{sec:effective}

\begin{figure}[t]
	\centering
	\includegraphics[width=\linewidth]{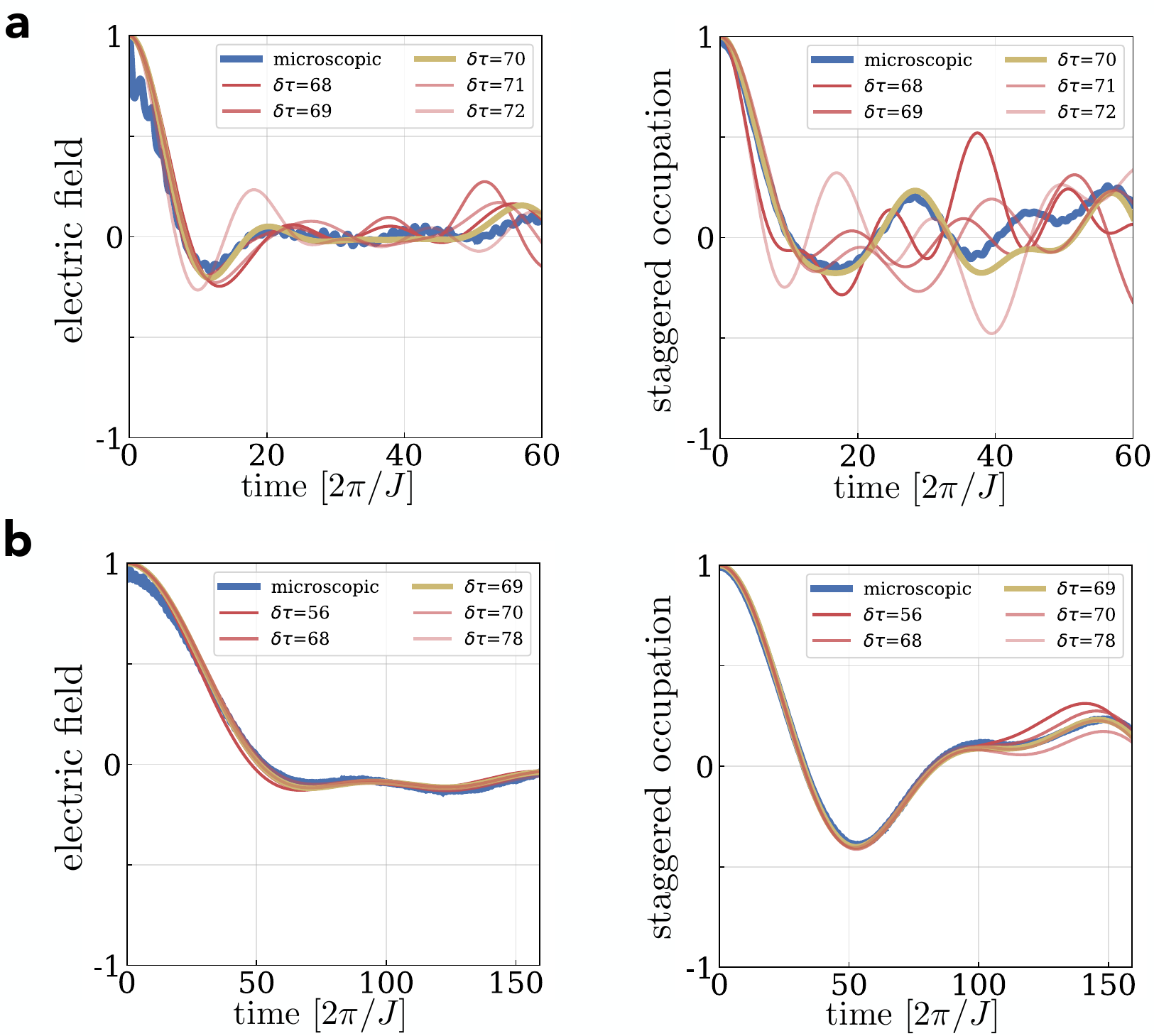}
    \caption{\textbf{Comparison of the effective and microscopic models for the $\mathrm{SU}(2)$ LGT with dynamical matter considered in Fig.~\ref{fig4}b.} \textbf{(a)} Rishon tunneling strength is set to $t=4t_m=0.15J$ as in Fig.~\ref{fig4}b. We find the best agreement for $\delta\tau=70/J$. \textbf{(b)} For a rishon tunneling strength of $t=4t_m=0.075J$, we find the best agreement for $\delta\tau=69/J$.}
\label{figEffDyn}
\end{figure}

\begin{figure*}[t!]
	\centering
	\includegraphics[width=\linewidth]{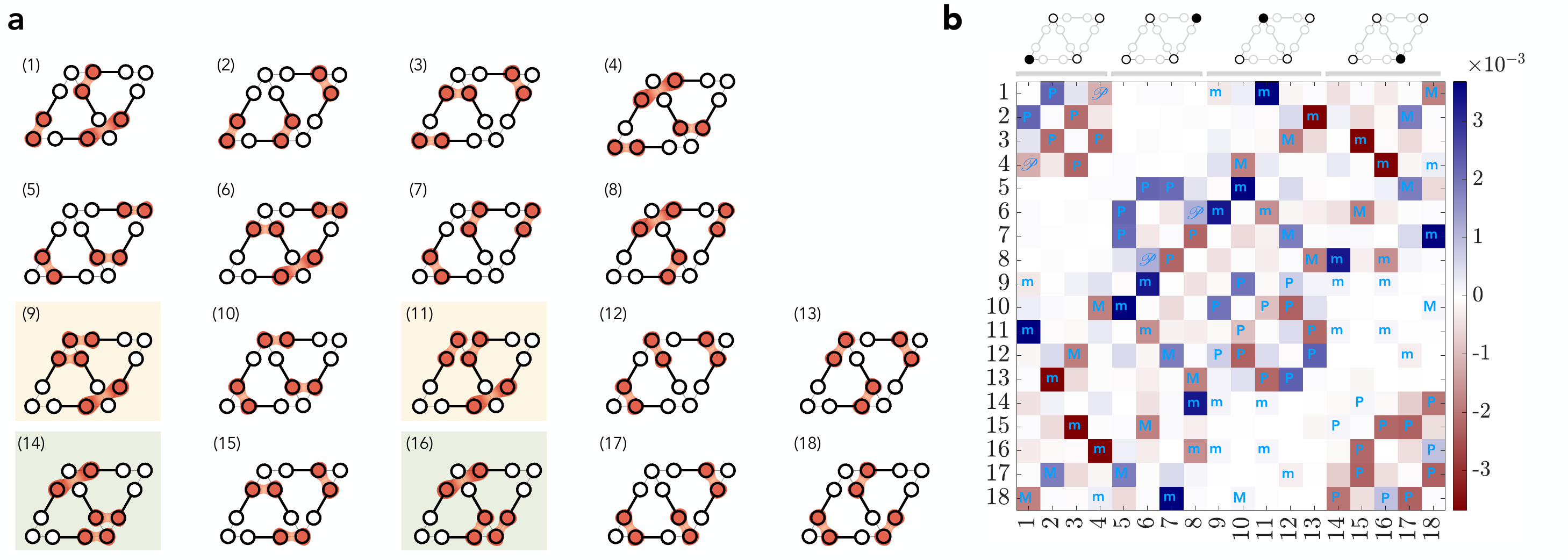}
    \caption{\textbf{Effective gauge theory with dynamical matter.} \textbf{(a)} The gauge-invariant basis of the $\mathrm{SU}(2)$ LGT with dynamical matter considered in Fig.~\ref{fig4}b. States with the same shade carry nonzero overlap with one another and are orthonormalized. \textbf{(b)} Matrix elements of the effective LGT as obtained numerically in units of $J$ from Eq.~\eqref{eq:PT} for $t=4t_m=0.075J$ and $\delta\tau=69/J$. Note that the basis states are ordered according to the location of the matter excitation, as indicated on top. Dominant matrix elements marked by `M' denote matter--rishon pair tunneling from one vertex to a neighboring vertex, while those marked with `m' denote matter--anti-rishon pair tunneling (minimal coupling) from a vertex to one of its neighbors. Dominant matrix elements denoting plaquette processes are either only $\mathrm{SU}(2)$ gauge-invariant, marked by `P', or $\mathrm{U}(2)$ gauge-invariant, marked by `$\mathscr{P}$'.}
\label{figEffMatrix}
\end{figure*}

\begin{figure}[t]
	\centering
	\includegraphics[width=\linewidth]{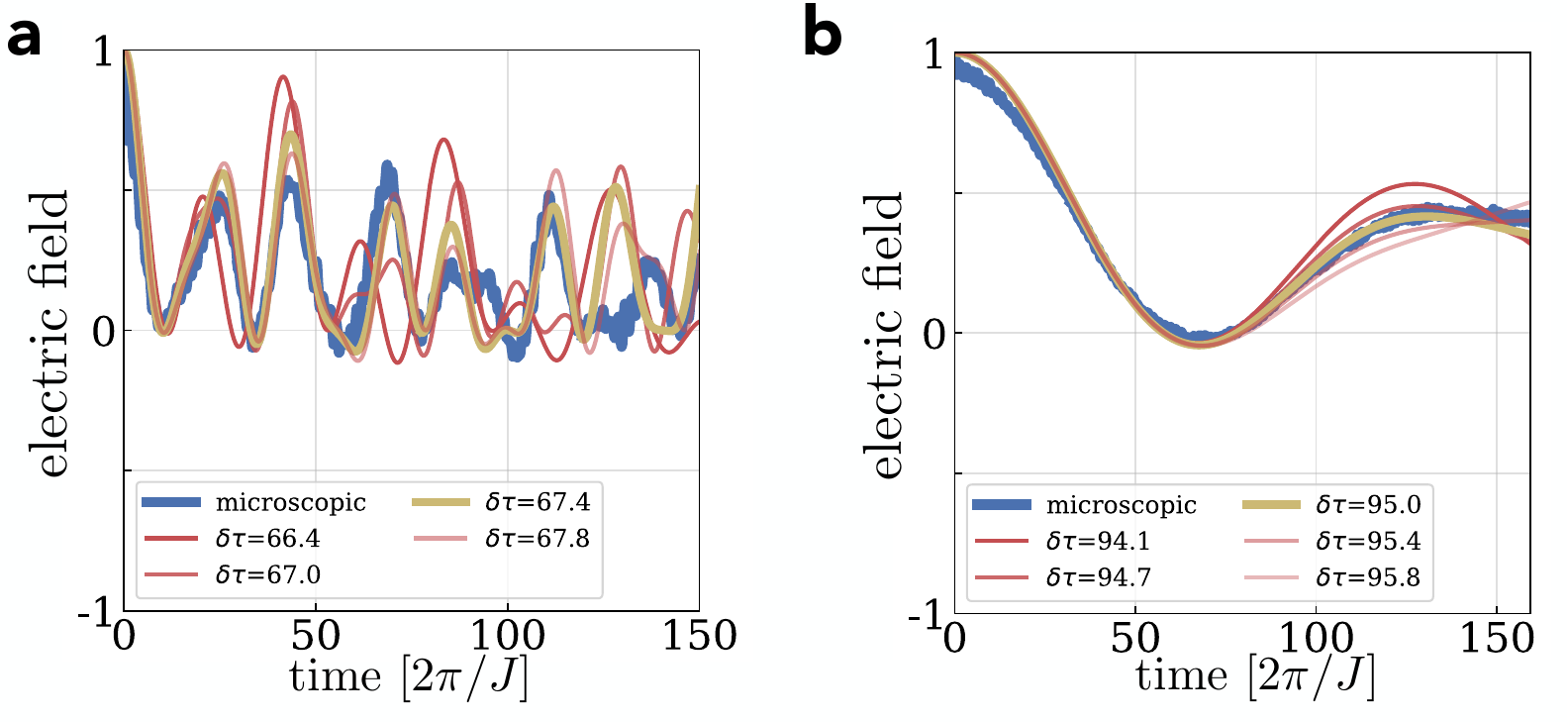}
    \caption{\textbf{Comparison of the effective and microscopic models for the pure $\mathrm{SU}(2)$ LGT considered in Fig.~\ref{fig4}a.} \textbf{(a)} Rishon tunneling strength is set to $t/J=0.075$ as in Fig.~\ref{fig4}a. We find the best agreement for $\delta\tau=67.4/J$. \textbf{(b)} For a rishon tunneling strength of $t/J=0.0375$, we find the best agreement for $\delta\tau=95/J$.}
\label{figEffDyn_Mstar}
\end{figure}

\begin{figure*}[t!]
	\centering
	\includegraphics[width=\linewidth]{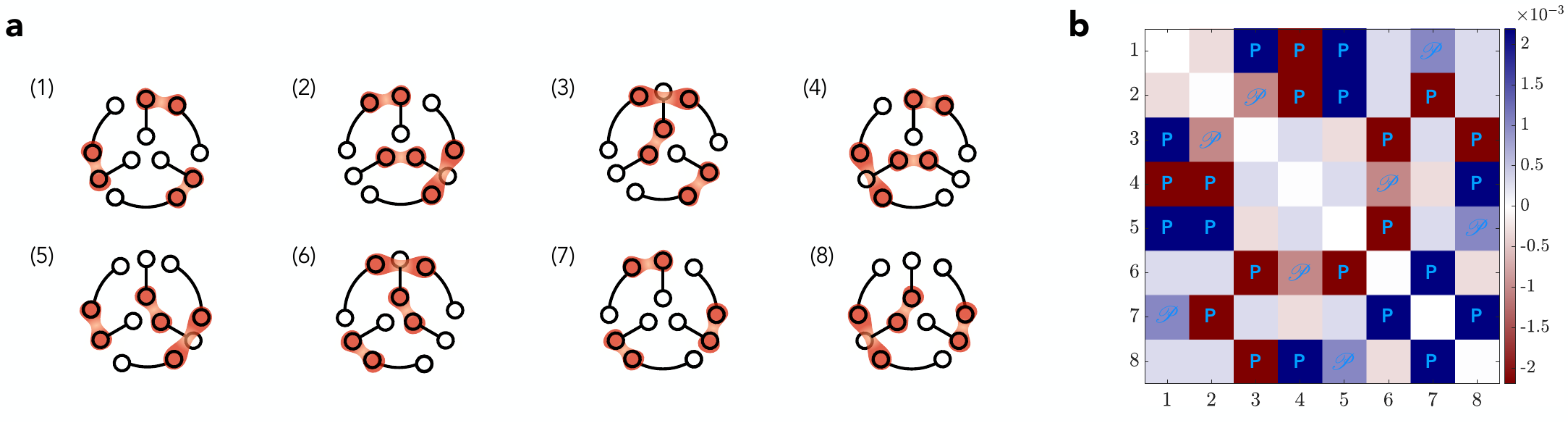}
    \caption{\textbf{Effective pure gauge theory.} \textbf{(a)} The gauge-invariant basis of the pure $\mathrm{SU}(2)$ LGT considered in Fig.~\ref{fig4}a. \textbf{(b)} Matrix elements of the effective LGT as obtained numerically in units of $J$ from Eq.~\eqref{eq:PT} for $t=0.0375J$ and $\delta\tau=95/J$. Dominant matrix elements denoting plaquette processes are either only $\mathrm{SU}(2)$ gauge-invariant, marked by `P', or $\mathrm{U}(2)$ gauge-invariant, marked by `$\mathscr{P}$'.}
\label{figEffMatrix_Mstar}
\end{figure*}

\begin{figure}[t]
	\centering
	\includegraphics[width=0.95\linewidth]{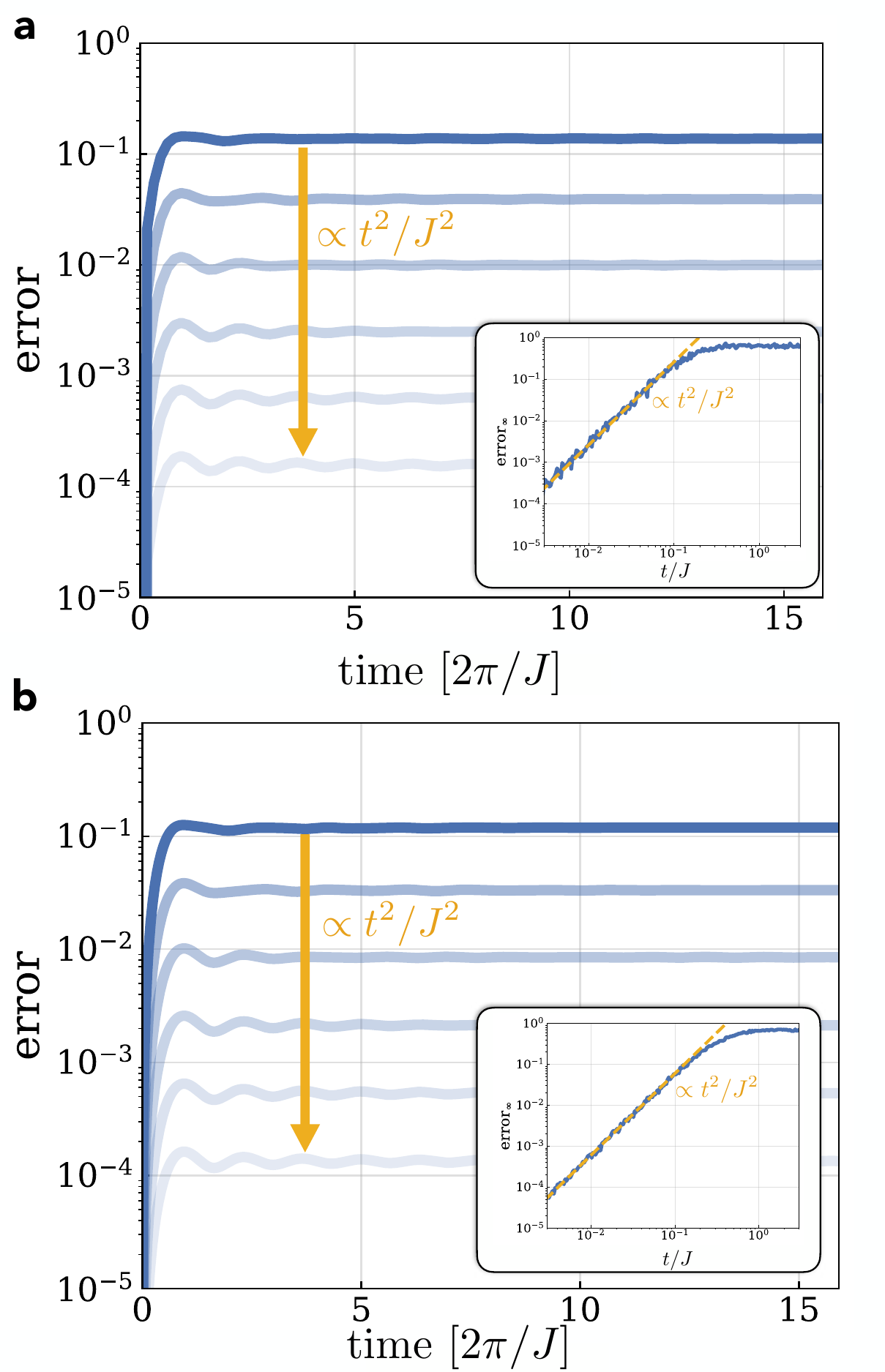}
    \caption{\textbf{Controlled gauge violation regime.} Using the same initial states as in Fig.~\ref{fig4}, we repeat the corresponding quenches for \textbf{(a)} the Mercedes star geometry at $t=0.075\times2^{-k}J$ and \textbf{(b)} the double-triangle geometry at $t=4t_m=0.15\times2^{-k}J$ with $k=0,1,\ldots,5$ (dark to bright blue). In all cases, we find that the gauge violation settles into a plateau of value $\propto t^2/J^2$ for all investigated evolution times, a hallmark of the controlled-violation regime \cite{Halimeh2020a}. In the insets, we show the ``infinite''-time gauge violation as a function of $t/J$. In ED, we chose various times $t\gtrsim10^4/J$ to extract this quantity. For sufficiently small $t/J$, stabilized gauge invariance is guaranteed.}
\label{figGaugeViolation}
\end{figure}

The \textit{top-down} approach introduced in the main text enables the precise engineering of the local symmetries of the emergent, low-energy effective model.
Before discussing quantitative numerical studies, we determine the resonant processes defined by the low-energy manifold of the protection term~$\hat{H}_J$, Eq.~\eqref{eq:GaugeProtection}, which yields the dominant terms of the effective model.

In Fig.~\ref{fig2}, we illustrate the approach on the experimentally feasible model with exactly \emph{one} rishon per link on the honeycomb lattice and~$N=2$.
We emphasize that the method applies for any~$N$, \emph{fixed} rishon number~$\mathcal{N}_L$ per link, coordination number, or particle statistics.

The most relevant terms that arise in the effective model are composed of intra- and inter-vertex processes, see Fig.~\ref{fig2}a.
While the intra-vertex processes conserve the color singlet constraint, they additionally conserve the total number of particles around a vertex, i.e., they have a local $\mathrm{SU}(N)\cross\mathrm{U}(1) \cong \mathrm{U}(N)$ symmetry.
The inter-vertex processes, on the other hand, conserve the color singlet constraint on every vertex but not the particle number; hence, they only generate $\mathrm{SU}(N)$~symmetric terms.

By constructing the effective model as described above, we can in principle obtain the form of the effective Hamiltonian with the typical minimal gauge-matter couplings and magnetic plaquette interactions.
In particular, two distinct types of effective interactions occur, see Fig.~\ref{fig2}b:
(i) Collective processes generated by intra-vertex terms only, e.g., the minimal gauge-matter coupling, are present in both $\mathrm{SU}(N)$ and $\mathrm{U}(N)$~LGTs.
(ii) Collective processes composed of intra- and inter-vertex terms, while maintaining the rishon number constraint~$\mathcal{N}_L$, only appear in $\mathrm{SU}(N)$~LGTs.

The amplitudes of these terms can have a functional dependence on the matter and rishon distribution, as well as on the particle statistics and can be computed perturbatively.
The leading minimal coupling and rishon-matter tunneling terms occur in second-order processes with strength~$\propto t\cdot t_m/J$; hence our proposal enables a feasible implementation of non-Abelian LGT with sizable effective interactions, see Fig.~\ref{fig4}.

In general, the leading-order contributions are generated by perturbatively coupling the energetically-resonant gauge-invariant states.
The protection term~\eqref{eq:GaugeProtection} enforces the color singlet constraint on each vertex, and in its simplest form in Eq.~\eqref{eq:GaugeProtection}, all singlets are energetically equally favoured.
On the other hand, by introducing disordered protection strengths,~$J \rightarrow J_\mathbf{r}$, transporting color singlets between vertices is energetically costly, which efficiently suppresses inter-vertex processes of type~(ii); therefore an additional local~$\mathrm{U}(1)$~symmetry is protected for a proper choice of $J_\mathbf{r}$.
In conclusion, our protection scheme allows for the controlled protection and dynamical simulation of both $\mathrm{SU}(N)$ and $\mathrm{U}(N)$~LGTs

Let us now turn to a concrete example, namely that of the $\mathrm{SU}(2)$ LGT with dynamical matter considered in Fig.~\ref{fig4}b, and understand the effective gauge theory arising from the employed microscopic Hamiltonian $\hat{H}=\hat{H}_J+\hat{H}_t+\hat{H}_{t_m}$. To achieve this, it is instructive to notice that
\begin{align}\label{eq:PT}
    \hat{\mathcal{P}}_Ge^{-i\hat{H}\delta\tau}\hat{\mathcal{P}}_G\approx e^{-i\hat{H}_\text{eff}\delta\tau},
\end{align}
where $\hat{\mathcal{P}}_G$ is the projector onto the gauge-invariant basis and $1/J\ll\delta\tau\ll J/(t\cdot t_m)$. One can then tune $\delta\tau$ to find the best agreement between the dynamics of local observables under the effective lattice gauge theory $\hat{H}_\text{eff}$ and that under the microscopic Hamiltonian $\hat{H}$. For $t=4t_m=0.15J$, which is the case of Fig.~\ref{fig4}b, such a comparison is shown in Fig.~\ref{figEffDyn}a, where $\delta\tau=70/J$ provides very good agreement. Choosing $t=4t_m=0.075J$, we find that the optimal value is $\delta\tau=69/J$. The corresponding effective gauge theory Hamiltonian in the latter case is illustrated in Fig.~\ref{figEffMatrix}. The matrix elements indicate dominant gauge-invariant minimal coupling and plaquette terms similar to those illustrated in Fig.~\ref{fig2}.

We note that this form of the matrix elements is qualitatively the same for all values of $t/J$ in the controlled-error regime where the infinite-time gauge violation is $\propto t^2/J^2$, see Fig.~\ref{figGaugeViolation}. So long as we are in that regime, we can always extract faithful gauge-theory physics from our quantum simulator \cite{Halimeh2020a,Halimeh2021gauge}. We have defined our gauge violation as $\sum_\mathbf{r}\langle\hat{\mathbf{G}}_\mathbf{r}^2\rangle/\sum_\mathbf{r}\langle\hat{\mathbf{G}}_\mathbf{r}^2\rangle_\infty$, where $\langle\cdot\rangle_\infty$ denotes the infinite-temperature state.

For completeness, we repeat the above analysis for the pure $\mathrm{SU}(2)$ LGT considered in Fig.~\ref{fig4}a without dynamical matter. The corresponding results for the dynamics and effective Hamiltonian matrix elements are shown in Figs.~\ref{figEffDyn_Mstar} and~\ref{figEffMatrix_Mstar}, respectively. Also here, we find that the electric-field dynamics due to the microscopic Hamiltonian $\hat{H}=\hat{H}_J+\hat{H}_t$ is well-reproduced by an effective gauge-invariant Hamiltonian for all investigated evolution times.

\section{Bosonic excitation tunneling scheme}\label{sec:bosonic}
In the main text, we have discussed a scheme in which two molecules form a link and the internal states of these molecules host exactly one (bosonic) rishon excitation.
Likewise, each matter site is occupied by a molecule. 
The internal rotational states of the molecules are used to encode the color charges of rishons and matter.
In particular, the three states~$\{\ket{\varnothing},\ket{ \downarrow}, \ket{\uparrow}\}$ are implemented in the rotational states $\{\ket{\mathcal{R}};\,\mathcal{R}=0,1,2\}$ of molecules; hence the tunneling of particles corresponds to an exchange of bosonic, molecular excitations.
Here, we want to fill in more details about the Floquet sequence, which is adapted from Ref.~\cite{Homeier2023}.

\begin{figure}[t]
	\centering
	\includegraphics[width=\linewidth]{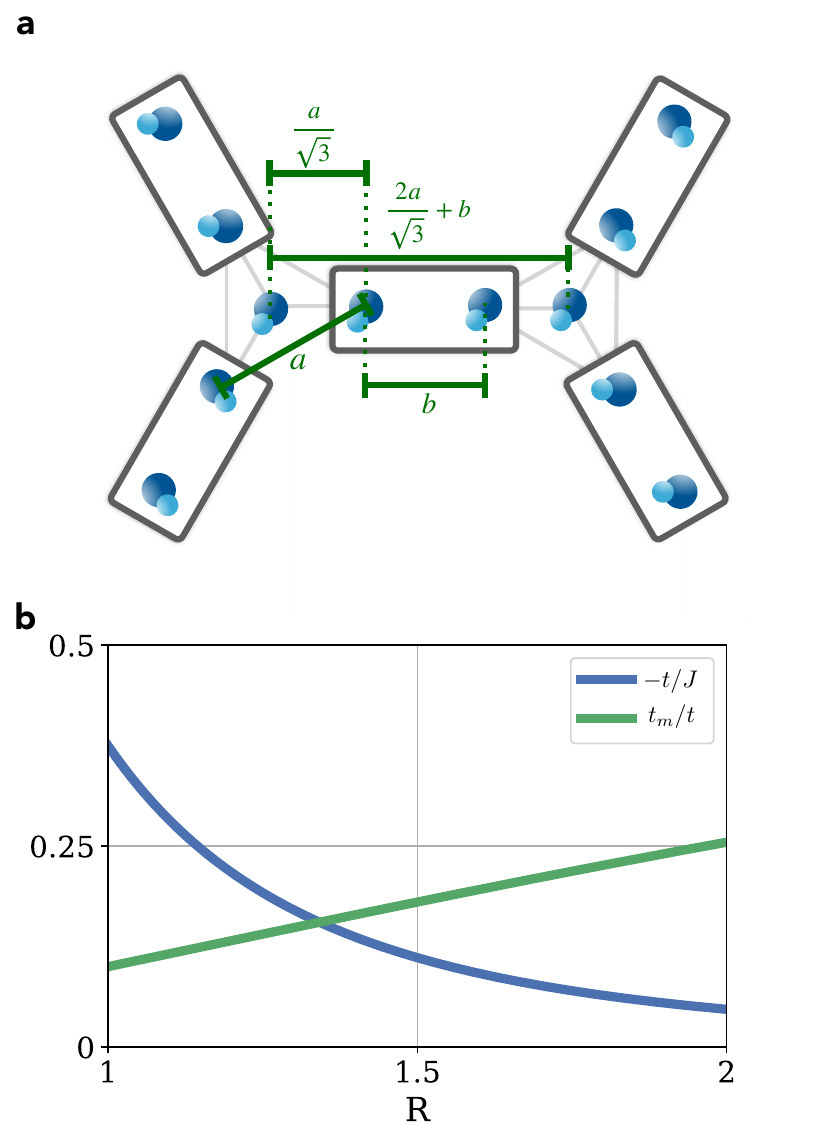}
    \caption{\textbf{Gauge theory with dynamical matter.} \textbf{(a)} To implement a gauge theory with dynamical matter, we need to (i) impose the gauge constraints and (ii) engineer the (weak) tunneling perturbation. This is realized by Floquet driving the resonant dipole interactions between molecules. Here, we show the geometry as in Fig.~\ref{fig3}b. The matter molecules at matter sites (center of triangle) are elevated out of the plane. \textbf{(b)} From the Floquet sequence illustrated in Fig.~\ref{fig3}b, we can determine the effective couplings with respect to~$R=b/a$. The Heisenberg interaction is given by~$J=\chi/3$, where~$\chi$ is the dipolar interaction at distance~$a$. }
\label{figSynHop}
\end{figure}

\textit{Floquet sequence.---}In the bosonic excitation tunneling scheme, we use three rotational states~$\ket{\mathcal{R}}$ of molecules that interact via resonant dipole-dipole interactions, i.e., flip-flop spin interactions between levels with~$\Delta \mathcal{R} = \pm 1$; here we assume that only the lowest two rotational states are dipole coupled.
The energy scale is determined by the dipole-dipole interaction~$\chi$ between adjacent rishon sites around a vertex, which have distance~$a$ as illustrated in Fig.~\ref{figSynHop}a.

To illustrate and calculate the effective Floquet Hamiltonian, we consider \textit{two} molecules labeled by sites~$i$ and~$j$ for simplicity.
The microscopic Hamiltonian is then given by
\begin{align}\nonumber
    \hat{H}_{\mathrm{mic}} =& \frac{\chi(r_{ij})}{2}\left( \hat{c}^\dagger_{i,\downarrow}\hat{c}_{j,\downarrow}  +\mathrm{H.c.} \right)\\\label{eq:supp_micHam}
\end{align}
which are the resonant dipole-dipole interactions between (polar) molecules at distance~$r_{ij}$.
We assume that the quantization axis is perpendicular to the plane of molecules.
Note that the first term, tunneling of $\downarrow$-excitations, can be suppressed by energetically detuning the $\ket{\mathcal{R}=0}$ state between the two molecules.

To determine the effective Floquet Hamiltonian from the sequence described in the main text and Fig.~\ref{fig3}b, we need to rotate and time evolve Eq.~\eqref{eq:supp_micHam} for time~$\tau_n$ accordingly.
For the nine intervals $n=1,\ldots,9$ shown in Fig.~\ref{fig3}b, we find the following resonant terms, i.e., terms that conserve the total number of rotational excitations,
\begin{subequations}
\begin{align}
    \hat{H}_1 =& \hat{H}_\mathrm{mic}, \\
    \hat{H}_2 =& \frac{\chi(r_{ij})}{2}\left(\hat{c}^\dagger_{i,\uparrow}\hat{c}_{j,\uparrow} + \mathrm{H.c.} \right)\\
    \hat{H}_3 =& \hat{H}_1\\
    \hat{H}_4 =& \chi(r_{ij})\left( \hat{S}^{x}_{i} \hat{S}^{x}_{j}  + \hat{S}^{y}_{i} \hat{S}^{y}_{j} \right), \\
    \hat{H}_5 =& \chi(r_{ij})\left( \hat{S}^{z}_{i} \hat{S}^{z}_{j}  + \hat{S}^{y}_{i} \hat{S}^{y}_{j} \right),\\
    \hat{H}_6 =& \hat{H}_4\\
    \hat{H}_7 =& \chi(r_{ij})\left( \hat{S}^{x}_{i} \hat{S}^{x}_{j}  + \hat{S}^{z}_{i} \hat{S}^{z}_{j} \right),\\
    \hat{H}_8 =& \hat{H}_4\\
    \hat{H}_9 =& \hat{H}_1
\end{align}
\end{subequations}
We neglect higher-order Floquet terms~$\mathcal{O}(t^2/T^2)$ with~$T=\sum_n \tau_n$, such that in total the first-order Floquet Hamiltonian reads
\begin{align}
\begin{split}
    \hat{H}_{\mathrm{Floquet}} = t\left( \hat{c}^\dagger_{i,\downarrow}\hat{c}_{j,\downarrow} + \hat{c}^\dagger_{i,\uparrow}\hat{c}_{j,\uparrow} \right) + J \hat{\mathbf{S}}_i \cdot \hat{\mathbf{S}}_j.
\end{split}
\end{align}
Note that we are considering a Hilbert space that does not contain any double occupancy~$\hat{c}^\dagger_{j,\downarrow}\hat{c}^\dagger_{j,\uparrow}\ket{0}$.
Here, we have chosen
\begin{subequations} \label{eq:taus_withoutV}
\begin{align} 
    \frac{\tau}{T} &= \left( 24\epsilon^t + 9 \right)^{-1} \\
    \tau_1 &= \tau_3 =\tau_9=4\epsilon^t\cdot\tau \\
    \tau_2 &= 12\epsilon^t\cdot\tau \\
    \tau_4 &=  \tau_6=\tau_8=\tau \\
    \tau_5 &= \tau_7=3\tau,
\end{align}
\end{subequations}  
This yields an effective Heisenberg interaction of strength~$\frac{J}{\chi} = \left(4\epsilon^t+\frac{3}{2}\right)^{-1}$ and corresponding tunneling amplitude of~$-t = \epsilon^t \cdot J$, using the sign convention as in Eq.~\eqref{eq:tunneling}.

\textit{Tweezer geometry.---}Now, we turn to the building block described in the main text in Fig.~\ref{fig3}b.
The relevant distances in the proposed geometry are illustrated in Fig.~\ref{figSynHop}a.
We assume an energetic detuning of the~$\ket{\mathcal{R}=0}$ state of molecules on adjacent rishon links as well as between rishon molecules and matter molecules.
This ensures the tunneling~$t_{ij}$ to be restricted to (i) within rishon links and (ii) between matter sites.
The detunings can in principle be controlled by, e.g., choosing appropriate magic angle conditions~\cite{Rosenband2018, Ni2018}.

Moreover, we set~$\chi(a) \equiv \chi$, where~$a$ is the experimentally smallest distance that allows to implement the geometry in Fig.~\ref{figSynHop}a.
Therefore,~$J =\frac{2}{3}\chi$ is the largest energy scale in the system and we can tune (perturbative) rishon tunneling~$t$ and matter tunneling~$t_m$ by the distance~$b \geq a$ shown in Fig.~\ref{figSynHop}a.
We find
\begin{subequations}
\begin{align}
    \frac{-t}{J} &= \frac{3}{8} \cdot \frac{1}{R^3}, \\
    t_m &= t \cdot \left( \frac{\sqrt{3}R}{\sqrt{3}R+2} \right)^{3}, \\
    R &:= \frac{b}{a} \geq 1.
\end{align}
\end{subequations}
The effective couplings are plotted in Fig.~\ref{figSynHop}b.

\section{Multi-rishon model: Fermionic superexchange}\label{sec:multi}
To realize models beyond one rishon per link and arbitrary~$N$, we propose a scheme based on fermionic superexchange.
The goal is to implement the protection Hamiltonian~\eqref{eq:HJ}, which requires attractive Hubbard interactions and antiferromagnetic $\mathrm{SU}(N)$-invariant Heisenberg interactions around a vertex.
Moreover, the rishon number constraint must be conserved.

\begin{figure}[t]
	\centering
	\includegraphics[width=\linewidth]{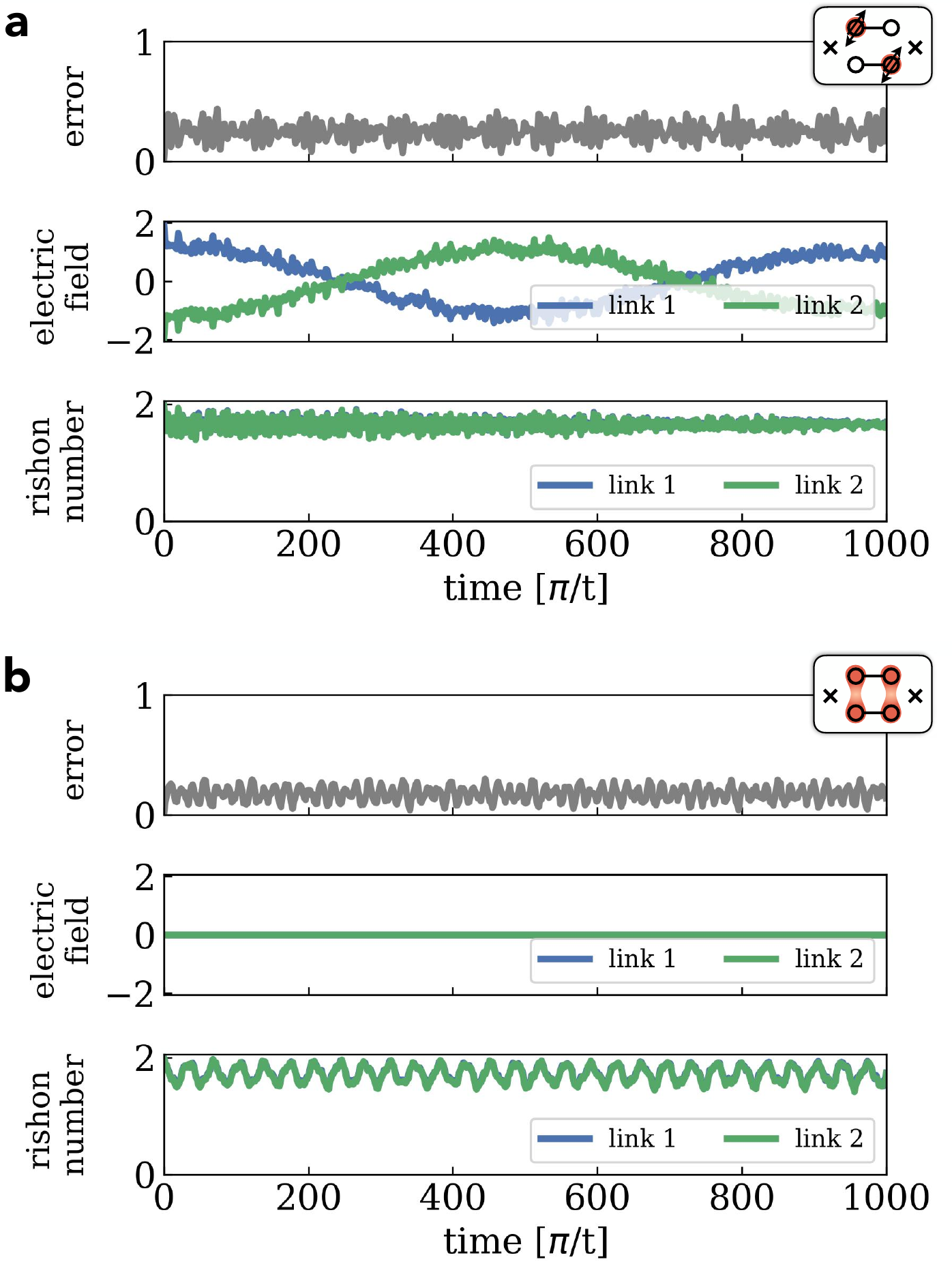}
    \caption{\textbf{Fermionic two rishon model.} We time evolve two different gauge invariant initial states in \textbf{(a)} and \textbf{(b)} under the proposed multi-rishon gauge protection scheme and the parameters discussed in Appendix~\ref{sec:multi-numerics}. The minimal setup is constituted by two links and initial states contain two rishons per link, see inset. We find that the gauge violation, normalized by the infinite temperature value, and the rishon number are efficiently protected in the presence of finite rishon tunneling perturbations~$t_r$. Further, the perturbation induces dynamics of the electric field shown in \textbf{(a)}. For the initial state in \textbf{(b)}, we find perfect destructive interference of the electric field dynamics.}
\label{figMultiNumerics}
\end{figure}

Hence, we propose to use ultracold fermionic, alkaline-earth atoms with weak $\mathrm{SU}(N)$-invariant, attractive Hubbard interactions~$U$ in an engineered optical potential landscape with $\lvert \delta\rvert,\lvert \Delta\rvert \gg \lvert t\rvert$, see Fig.~\ref{fig5}.
The superexchange process is mediated through an auxiliary site on each vertex.
In contrast to the scheme for molecules, the AFM interactions are not implemented via long-range dipolar interactions and thus the interaction strength is not geometrically constrained; hence the equal strength interactions can -- in principle -- be obtained for any arbitrary geometry by coupling rishon and matter sites to an auxiliary site.

The AFM Heisenberg interactions around a vertex arise from fourth-order processes, which we obtain by performing a Schrieffer-Wolff transformation.
We find that the effective fourth-order vertex terms of the setup shown in Fig.~\ref{fig5} are given by
\begin{align}
    \hat{H}^{(4)}_\mathrm{eff} = \hat{H}_{\mathrm{AFM}} + \hat{H}_\text{nn},
\end{align}
with
\begin{align} \label{eq:SuperexchangeH}
    \hat{H}_{\mathrm{AFM}} &= \sum_{\mathbf{r},a} \sum_{b<a} J\left[ \hat{n}_{(\mathbf{r},a)},\hat{n}_{(\mathbf{r},b)}\right] \hat{\mathbf{S}}_{\mathbf{r},a} \cdot \hat{\mathbf{S}}_{\mathbf{r},b}, \\
    \hat{H}_\text{nn} &= \sum_{\mathbf{r},a} \sum_{b<a} V\left[ \hat{n}_{(\mathbf{r},a)},\hat{n}_{(\mathbf{r},b)}\right]  \hat{n}_{(\mathbf{r},a)}\hat{n}_{(\mathbf{r},b)}.
\end{align}
The amplitudes $J$ and~$V$ are weakly density-dependent because the energy difference to the virtual states depends on the number of rishons at a given rishon site through the interaction~$U$.
Nevertheless, all terms in Eq.~\eqref{eq:SuperexchangeH} are $\mathrm{SU}(N)$ invariant.

The amplitudes are related by
\begin{align}
    V\left[ \hat{n}_{(\mathbf{r},a)},\hat{n}_{(\mathbf{r},b)}\right] = \frac{1}{2}J\left[ \hat{n}_{(\mathbf{r},a)},\hat{n}_{(\mathbf{r},b)}\right]\left( \frac{1}{N} - 1\right),
\end{align}
which gives the well-known $V= -J/4$ density-density term of the $t$\,--\,$J$ model for $N=2$~\cite{Auerbach_book}.
The strength~$J$ scales as~$\propto t^4/\Delta^3$, and we require that
\begin{align} \label{eq:GausslawCondition}
    J = -2NU/(1+N)
\end{align} for the on-site Hubbard interaction, Eq.~\eqref{eq:HJ}.
Therefore, we require the Hubbard interaction to be small~$\lvert U\rvert \ll \lvert \delta\rvert,\lvert \Delta\rvert$.
In this limit, we can expand the AFM interaction~$J\left[ \hat{n}_{(\mathbf{r},a)},\hat{n}_{(\mathbf{r},b)}\right]$ in the small $\lvert U\rvert$ and obtain the density-independent coupling strength
\begin{align} \label{eq:superexchangeAFM}
    J = \dfrac{16t^4 \left( 36 \Delta^2 -5 \delta^2 \right) }{\left( \delta^2 -4\Delta^2 \right)^3} U + \mathcal{O}\left( U^2 \right).
\end{align}
The desired gauge protection term can be constructed by choosing a set of parameters~$(t, U, \delta)$ and finding~$\Delta$ such that Eq.~\eqref{eq:GausslawCondition} is fulfilled.

\section{Multi-rishon model: Numerical simulation}\label{sec:multi-numerics}

We test our proposed superexchange gauge protection scheme for pure multi-rishon ${\rm SU}(2)$~models using exact diagonalization on a minimal example. To this end, we simulate a system with two vertices connected by two links, where each link contains two fermionic rishons. The rishon sites adjacent to a vertex interact via a superexchange process mediated by virtually tunneling on the energetically detuned auxiliary site, see Fig.~\ref{fig5} and Appendix~\ref{sec:multi}. We choose the parameters according to Eqs.~\eqref{eq:GausslawCondition} and \eqref{eq:superexchangeAFM}, i.e., we require
\begin{equation}
    -\frac{4}{3}U = \dfrac{16t^4 \left( 36 \Delta^2 -5 \delta^2 \right) }{\left( \delta^2 -4\Delta^2 \right)^3}U,
\end{equation}
which is fulfilled for~$t=1$, $\Delta/t \approx -10.35$, $\delta/t=20$. Further, we set $U/t=-0.25$ yielding a protection strength of~$J \approx 0.32 t$.
The tunneling amplitude~$t$ denotes hopping within vertices and we introduce the tunneling $t_r=J/8$ as the perturbation on rishon links.

Using exact numerical simulations, we time evolve gauge invariant initial states and evaluate (i) the Gauss' law error normalized by the infinite temperature gauge violation, (ii) the electric field defined as the imbalance of the rishon number on a link, and (iii) the total rishon number per link. In Fig.~\ref{figMultiNumerics}, we show the results for two different gauge invariant initial states and find that the gauge breaking errors are controlled and remain at about $10-15\%$ for the given set of parameters. Moreover, the perturbation induces gauge invariant effective dynamics indicated by the observed dynamics in the electric field, see Fig.~\ref{figMultiNumerics}(a). For the initial state in Fig.~\ref{figMultiNumerics}(b), we find exact destructive interference in the electric field; nevertheless the oscillations in other quantities show that our proposed gauge protection scheme is able to efficiently suppresses gauge violating terms, e.g., singlet-triplet oscillations induced by second-order superexchange~$\propto 4t_r^2/U$ on links.

\end{document}